\documentclass[pdflatex]{revtex4}

\usepackage{latexsym}
\usepackage{amsmath,amssymb,bm} 
\usepackage{comment,ulem} 
\usepackage[pdftex]{graphicx} 
\usepackage[pdftex, colorlinks=true, linkcolor=blue]{hyperref}

\makeatletter

\newcommand{\cket}[1]{| #1 \rangle} 
\newcommand{\bra}[1]{\langle #1 |} 
\newcommand{\ImUnit}{\mathrm{i}} 

\begin{document}

\title{Quantum teleportation in vacuum using only Unruh-DeWitt detectors} 

\author{Jun-ichirou Koga} 
\affiliation{Research Institute for Science and Engineering, 
Waseda University, Shinjuku-ku, Tokyo 169-8555, Japan} 
\email{koga@waseda.jp}
 
\author{Gen Kimura} 
\affiliation{College of Systems Engineering and Science, 
Shibaura Institute of Technology, Saitama 330-8570, Japan} 
\email{gen@shibaura-it.ac.jp}

\author{Kengo Maeda} 
\affiliation{Faculty of Engineering, 
Shibaura Institute of Technology, Saitama 330-8570, Japan} 
\email{maeda302@sic.shibaura-it.ac.jp}

\begin{abstract} 
We consider entanglement extraction into two two-level Unruh-DeWitt detectors from a vacuum of a neutral massless quantum scalar field in a four-dimensional spacetime, 	
where the general monopole coupling to the scalar field is assumed. 
Based on the reduced density matrix of the two detectors derived within the perturbation theory, 
we show that {\it the single copy} of the entangled pair of the detectors can be utilized in quantum teleportation even when the detectors are separated acausally, while we observe no violation of the Bell-CHSH inequality. 
In the case of the Minkowski vacuum, in particular, 
we find that entanglement usable in quantum teleportation is extracted due to the special relativistic effect 
when the detectors are in a relative inertial motion, while it is not  
when they are comoving inertially and the switching of the detectors is executed adiabatically at infinite past and future.    
\end{abstract}

\maketitle 

\section{Introduction} 
\label{sec:Introduction} 
\eqnum{0} 

Quantum entanglement in the relativistic quantum field theory is a developing field of research. 
From the point of view of theoretical physics, in particular, 
the information loss paradox in black hole physics (see, e.g., Refs. \cite{Hawking76,AlmheiriMPSS13}) and 
the entanglement entropy in anti-de Sitter/conformal field theory correspondence \cite{RyuTakayanagi06} have attracted  
much attention recently. 

It has been shown by Summers and Werner \cite{SummersWerner85,SummersWerner87a,SummersWerner87b,SummersWerner87c} that, with {\it suitable} observables in local spacetime regions, the Bell--Clauser-Holt-Shimony-Horne (CHSH) inequality is maximally violated in a vacuum of any quantum field theory, 
which has led to the observation that the vacuum intrinsically contains the ``non-local '' correlations, hence also entanglement, that cannot be explained from a local realistic view \cite{ref:Bell,ref:CHSH,ref:Werner}. Then, much effort has been made to analyze entanglement extraction from a vacuum with detectors so that it can be useful for several quantum information processing methods. 
(For entanglement extraction from the Minkowski vacuum, see, e.g., \cite{AlsingMilburn03,Fuentes-SchullerMann05,LInCH08,LinHu09,LinCH15,Reznik03,ReznikRS05,Braun05,LeonSabin09a,LeonSabin09b,LeonSabin08,LeonSabin09c,ClicheKempf10,SaltonMM15,Pozas-KerstjensMartinMartinez15,MartinMartinezST16,Pozas-KerstjensMartinMartinez16,Pozas-KerstjensLM17,SimidzijaMartin-Martinez17a,SimidzijaMartin-Martinez17b,NambuOhsumi09}.)  
In particular, following the pioneering paper by Reznik \cite{Reznik03}, two-level Unruh-DeWitt detectors \cite{Unruh76,DeWitt79,BirrellDavies82,LoukoSatz08} are frequently employed to analyze entanglement extracted from a vacuum, where we can apply the well-established results in quantum information theory for qubits. In such extraction of entanglement, which is called harvesting, Unruh-DeWitt detectors are assumed 
to be localized not only spatially but also temporally. 
In particular, Reznik \cite{Reznik03} considered Unruh-DeWitt detectors that interact with a quantum field for a strictly finite period 
so that the future light cones of the detectors do not intersect within the period of interaction.  
It thus has revealed  
that entanglement is generated between detectors even if they are separated acausally, i.e., located at causally disconnected regions, 
and has corroborated the result by Summers and Werner  \cite{SummersWerner85,SummersWerner87a,SummersWerner87b,SummersWerner87c}.   
Since the maximally entangled state can be distilled from an ensemble of any two-qubit entangled states \cite{HorodeckiHH97}, one can in principle make use of the extracted entanglement for some information processes such as quantum teleportation. Notice, however, such distillation requires the preparation of infinitely many copies of vacua and detectors, which may be rather unrealistic.  
Therefore, it is still meaningful to ask whether the single copy of the entangled state has potential abilities, especially in the case where the detectors are separated acausally. 
The first purpose in this paper is then to understand in a general context the usability of the entanglement between the Unruh-DeWitt detectors coupled to a neutral massless quantum scalar field through a monopole coupling, i.e., without internal structures of a detector. 
In particular, we will show that, although the entanglement does not violate the Bell-CHSH inequality, a quantum teleportation with the use of the single copy of the entangled Unruh-DeWitt detectors is still possible. 
To see this, we will assume neither the geometry of a spacetime, the form of the monopole coupling, the worldlines of the detectors, 
nor the switching functions of the detectors. 

The notion of the Unruh-DeWitt detector has been introduced as a theoretical device that probes the nature of a quantum state. In the Minkowski vacuum, in particular, it detects no excitation when it is carried by an inertial observer, and its spectrum is thermal if it is carried by a uniformly accelerated observer. The latter particularly has corroborated the observation that the Minkowski vacuum looks 
to be a thermal bath for a uniformly accelerated observer, which is the so-called Unruh effect \cite{Unruh76}. These results follow when one switches on and off the detector adiabatically at infinite past and future, which is implicitly assumed in the textbooks \cite{DeWitt79,BirrellDavies82}, 
and computes the excitation probability 
by considering practically detectors that interact with a quantum field for infinitely long time, and thus globally in time. 
As for entanglement, one might expect that a sufficiently long interaction time will naturally enable entanglement extraction. 
This is not necessarily the case, however. In the realistic model \cite{LeonSabin09a,LeonSabin09b,LeonSabin08,LeonSabin09c}, for example, each of a pair of atoms with the electric dipole $\boldsymbol{d}$ is coupled with 
the electric field $\boldsymbol{D}$, whose interaction Hamiltonian is given as $H_I = - ( 1 / \varepsilon_0 ) \, \boldsymbol{d \cdot D}$, 
and the initial quantum state of the electromagnetic field is set to be the vacuum, i.e., 
without applying any external electric field. 
If the initial states of the atoms are prepared to be the ground state, it might be conceivable 
that the uncertainty relation in time and energy 
suppresses quantum fluctuation after sufficiently long time, the energy conservation being recovered, 
and the whole system returns back to the ground state. 
Then, the interaction might be expected as ineffective. 
In fact, as we will show in this paper, when two Unruh-DeWitt detectors are comoving inertially and interact with a quantum field for 
infinitely long time, entanglement is {\it not} generated between the Unruh-DeWitt detectors.  
Thus, the second purpose of this paper is to demonstrate it explicitly and further explore non-comoving inertial motions of Unruh-DeWitt detectors that interact with a quantum field for infinitely long time. 

We will thus consider in this paper two two-level Unruh-DeWitt detectors coupled to a neutral massless quantum scalar field with 
a general monopole coupling, 
where the initial states of the detectors are prepared to be the ground state and the initial state of the scalar field 
is set to be a vacuum. The model is presented in Sec.\ref{sec:DensityMatrix}, where we will also 
derive the reduced density matrix of the two Unruh-DeWitt detectors in an arbitrary four-dimensional spacetime in the perturbation theory, without assuming particular forms of the worldlines of the detectors or the switching functions. 
Based on the reduced density matrix, in Sec.\ref{sec:Measures}, we will compute several entanglement measures, which include 
the bounds on the distillable entanglement, the entanglement cost, and the Bell-CHSH inequality. In particular, we will see the supremacy of the extracted entanglement in quantum teleportation with the single entangled pair, compared to the teleportation only via classical channels. 
In Sec.\ref{sec:MinkowskiEn}, we will turn to the issue of entanglement generated between inertial Unruh-DeWitt detectors with adiabatic switching. To perform explicit computation, 
we need to specify a spacetime and a vacuum, 
and thus we will focus on the Minkowski vacuum. 
We will consider not only the case of infinitely long interaction with switching executed implicitly at infinite past and future, 
but also the case where the effect of switching is taken into account. 
We will conclude and discuss in Sec.\ref{sec:discussion}, which includes a preliminary result for the case where 
one of the detectors runs with uniform acceleration. 
Throughout this paper, we adopt natural units $c = \hbar = 1$.

\section{Reduced density matrix of two detectors} 
\label{sec:DensityMatrix} 

We consider two two-level Unruh-DeWitt detectors, $A$ carried by Alice and $B$ by Bob, 
with the discrete energy eigenvalues $E_n^{(A)}$ and $E_n^{(B)}$, respectively, where $n = 0 , 1$, which are thus considered as two qubits. 
The excitation energy $\Delta E^{(I)} \equiv E_1^{(I)} - E_0^{(I)}$ is assumed to be positive $\Delta E^{(I)} > 0$, where the index $I$ stands for $A$ or $B$, here and hereafter.
We denote the coordinate variables of the worldlines of the detectors with a bar as  
$\bar{x}^{\mu}_I(\tau_I)$, where $\tau_I$ is the proper time of the detector $I$. 
These detectors are assumed to be coupled with a neutral massless quantum scalar field $\phi(x)$,  
with the coupling being governed by the  perturbation action   
\begin{equation}
\mathcal{S}_{{\rm int}} 
= \int c \, \chi_A(\tau_A) \, m_A(\tau_A) \, \phi(\bar{x}_A) d \tau_A 
+ \int c \, \chi_B(\tau_B) \, m_B(\tau_B) \, \phi(\bar{x}_B) d \tau_B , 
\label{eqn:IntAction} 
\end{equation}
where $c$ is the coupling constant, 
$m_I(\tau_I)$ is the monopole operator of the detector $I$, which commutes with that of the other detector and with the scalar field 
$\phi\big(\bar{x}_I(\tau_I)\big)$. 
Note that we do not assume any particular form of the monopole coupling, guaranteeing the generality of the following discussion.  
The switching function $\chi_I(\tau_I)$ describes how the coupling between the detectors and the scalar field is implemented as a function of the proper time $\tau_I$. 

We choose the initial quantum state $\cket{\mathrm{in}}$ of the whole system at infinite past as 
\begin{equation}
\cket{\mathrm{in}} = \cket{0} \cket{E_0^{(A)}} \cket{E_0^{(B)}} , 
\label{eqn:IniWholeState}
\end{equation}
where $\cket{0}$ is the vacuum of the scalar field and $\cket{E_n^{(I)}}$ is the $n$ th state of the detector $I$. 
Then, the quantum state in the asymptotic future is given by 
\begin{equation} 
\cket{\mathrm{out}} = T \, e^{\ImUnit \, \mathcal{S}_{{\rm int}} 
} \, \cket{\mathrm{in}} , 
\label{eqn:OutState} 
\end{equation} 
where $T$ stands for time ordering. 
Since we are interested in the state of the two Unruh-DeWitt detectors, we trace out the degrees of freedom of the scalar field $\phi$. 
As is shown in Appendix \ref{sec:ExplicitMatrixElements}, the reduced density matrix $\rho_{A B}$ of the two detectors is then derived from the perturbation theory as 
\begin{equation} 
\rho_{AB} \equiv \mathrm{Tr}_{\phi} \cket{\mathrm{out}} \bra{\mathrm{out}} 
= \begin{pmatrix} 
0 & 0 & 0 & c^2 \: \mathcal{E} \\ 
0 & c^2 \: \mathcal{P}_A & c^2 \: \mathcal{P}_{AB} & c^2 \: \mathcal{W}_A \\ 
0 & c^2 \: \mathcal{P}_{AB}^* & c^2 \: \mathcal{P}_B & c^2 \: \mathcal{W}_B \\ 
c^2 \: \mathcal{E}^* & c^2 \: \mathcal{W}_A^* & c^2 \: \mathcal{W}_B^* & 
1 - c^2 \big( \mathcal{P}_A + \mathcal{P}_B \big) \end{pmatrix} + O(c^4) , 
\label{eqn:RedDensityMatrix} 
\end{equation} 
in the bases of $\{ \cket{E_1^{(A)}} \cket{E_1^{(B)}} , \, 
\cket{E_1^{(A)}} \cket{E_0^{(B)}} , \, \cket{E_0^{(A)}} \cket{E_1^{(B)}} , \, \cket{E_0^{(A)}} \cket{E_0^{(B)}} \}$, 
where $*$ stands for the complex conjugate.  
Among the matrix elements in Eq. (\ref{eqn:RedDensityMatrix}), $\mathcal{P}_I$ and $\mathcal{E}$ are given by 
\begin{equation} 
\mathcal{P}_I 
=  \left| \bra{E_{1}^{(I)}} \, m_I(0) \, \cket{E_0^{(I)}} \right|^2 \, \mathcal{I}_I , \quad 
\mathcal{E} 
= \bra{E_{1}^{(B)}} m_B(0) \cket{E_0^{(B)}} \, \bra{E_{1}^{(A)}} m_A(0) \cket{E_0^{(A)}} \, 
\mathcal{I}_E , 
\label{eqn:IntFactorsPE} 
\end{equation} 
where $\mathcal{I}_I$ and $\mathcal{I}_E$ are defined as 
\begin{align}  & 
\mathcal{I}_I \equiv 
\int_{- \infty}^{\infty} d \tau'_I \; \int_{- \infty}^{\infty} d \tau_I \; \chi_I(\tau'_I) \: \chi_I(\tau_I) \: 
e^{\ImUnit \, \Delta E^{(I)} \left( \tau_I -  \tau'_I \right)} \: G_W(\bar{x}'_I , \bar{x}_I) , 
\label{eqn:RedMatElPI} \\ & 
\mathcal{I}_E \equiv
- \:  \ImUnit \, \int_{- \infty}^{\infty} d \tau_B \, \int_{- \infty}^{\infty} d \tau'_A \:  \chi_B(\tau_B) \, \chi_A(\tau'_A) \, 
e^{\ImUnit \, \Delta E^{(B)} \tau_B} e^{\ImUnit \, \Delta E^{(A)} \tau'_A} \: G_F(\bar{x}_B , \bar{x}'_A) ,   
\label{eqn:RedMatElE1}
\end{align} 
and $G_W(x , x')$ and $G_F(x , x')$ are the Wightman function and the Feynman propagator 
defined as 
\begin{align} & 
G_W(x , x') \equiv \bra{0} \, \phi(x) \, \phi(x') \, \cket{0} , 
\label{eqn:WightmanNSDefGen} \\ & 
G_F(x , x') \equiv - \, \ImUnit \, \bra{0} \, T \phi(x) \, \phi(x') \, \cket{0} .  
\label{eqn:FeynmanNSDefGen} 
\end{align} 
We emphasize here that $\mathcal{P}_I$ is nothing but the excitation probability of the detector $I$ from the ground state to the excited state, and thus we have 
\begin{equation} 
\mathcal{P}_I \geq 0 . 
\label{eqn:PositivePs} 
\end{equation}  
We note also that Eq. (\ref{eqn:RedMatElE1}) is rewritten in terms of the Wightman function and the retarded Green function, 
as Eq. (\ref{eqn:CalEDefApp2}) in Appendix \ref{sec:ExplicitMatrixElements}, by using the relation among the Green functions (\ref{eqn:RelRetFeyWight}). 
The forms of other elements in Eq.~\eqref{eqn:RedDensityMatrix} are given in Appendix \ref{sec:ExplicitMatrixElements}.
Note that the density matrix \eqref{eqn:RedDensityMatrix} generalizes those derived in \cite{Reznik03,ReznikRS05,ClicheKempf10,SaltonMM15,MartinMartinezST16,NambuOhsumi09}  
in the sense that ${\cal W}_I$ appears in our case. Interestingly, we will see that ${\cal W}_I$ does not play any role in this paper, similarly to the case where the initial state of the scalar field is set to the coherent state and the detectors have spacial extension but with a restricted form of the monopole coupling \cite{SimidzijaMartin-Martinez17a,SimidzijaMartin-Martinez17b}. 

As we describe in Appendix \ref{sec:ExplicitMatrixElements}, the reduced density matrix (\ref{eqn:RedDensityMatrix}) is derived by employing only the relations between the Green functions derived from 
Eqs.(\ref{eqn:WightmanNSDefGen}) and (\ref{eqn:FeynmanNSDefGen}), along with the condition that 
the worldlines of the detectors are causal, i.e., the time coordinate $t(\tau)$ of a single timelike worldline is 
a monotonically increasing function of the proper time $\tau$ and hence 
\begin{equation} 
\Theta(\tau - \tau') \, \Theta(t(\tau) - t(\tau')) 
= \Theta(\tau - \tau') , \quad 
\Theta(\tau - \tau') \, \Theta(t(\tau') - t(\tau)) = 0 , 
\label{eqn:Causality} 
\end{equation} 
where $\tau$ and $\tau'$ are proper times along the single timelike worldline, 
and thus we see that the reduced density matrix given by Eq. (\ref{eqn:RedDensityMatrix}) is valid 
for arbitrary timelike worldlines of the detectors in an arbitrary spacetime, 
once a vacuum $\cket{0}$ is well-defined, not necessarily uniquely. 

The eigenvalues of the reduced density matrix (\ref{eqn:RedDensityMatrix}) are derived by using Eq. (\ref{eqn:EigenValuesRhoS}) in Appendix \ref{sec:Eigenvalues} as 
\begin{equation} 
1 + O(c^2) , \quad \dfrac{c^2}{2} \left[ \mathcal{P}_A + \mathcal{P}_B 
\pm \sqrt{\left( \mathcal{P}_A - \mathcal{P}_B \right)^2 + 4 \left| \mathcal{P}_{AB} \right|^2} \right] + O(c^4) , \quad 
c^4 \left( \mathcal{X} - \left| \mathcal{E} \right|^2 \right) + O(c^6) , 
\label{eqn:EigenVRho} 
\end{equation} 
where $\mathcal{X}$ arises in the contribution of order of $c^4$ as 
\begin{equation} 
c^4 \, \mathcal{X} \equiv \bra{E_1^{(A)}} \bra{E_1^{(B)}} \rho_{AB} \cket{E_1^{(A)}} \cket{E_1^{(B)}} ,  
\label{eqn:CalXDef} 
\end{equation}  
From the positivity of the density matrix $\rho_{AB}$ and Eq. (\ref{eqn:PositivePs}), we have  
\begin{equation}
\mathcal{P}_A \, \mathcal{P}_B \geq \left| \mathcal{P}_{AB} \right|^2 , \qquad \mathrm{and} \qquad \mathcal{X} \geq \left| \mathcal{E} \right|^2 . 
\label{eqn:PositivityRho} 
\end{equation} 
The reduced density matrix $\rho_B$ of the detector $B$ is obtained by further 
tracing out over the states of the detector $A$, as 
\begin{equation}
\rho_B \equiv \mathrm{Tr}_A \rho_{AB} = \begin{pmatrix} c^2 \: \mathcal{P}_B & c^2 \: \mathcal{W}_B \\ 
c^2 \: \mathcal{W}_B^* & 1 - c^2 \mathcal{P}_B \end{pmatrix} ,  
\end{equation} 
whose eigenvalues are derived as  
\begin{equation}
1 - c^2 \: \mathcal{P}_B + O(c^4) , \quad 
c^2 \: \mathcal{P}_B + O(c^4) . 
\label{eqn:EigenVRhoB} 
\end{equation}
\section{Properties of extracted entanglement} 
\label{sec:Measures} 

Based on the reduced density matrix (\ref{eqn:RedDensityMatrix}) of the two Unruh-DeWitt detectors, we here 
consider the possibility of entanglement extraction and its general properties by computing entanglement measures. 

The necessary and sufficient condition for a two-qubit system to be entangled, as two two-level Unruh-DeWitt detectors, is given by the famous positive partial transpose (PPT) criterion \cite{peres,HorodeckiHH96}: 
a two-qubit state $\rho_{AB}$ is entangled if and only if its partial transpose has negative eigenvalues. 
The partial transpose $\rho_{AB}^{T_A}$ of $\rho_{AB}$ with respect to the detector $A$ is given as 
\begin{equation}
\rho_{AB}^{T_A} = \begin{pmatrix} 
0 & 0 & 0 & c^2 \: \mathcal{P}_{AB}^* \\ 
0 & c^2 \: \mathcal{P}_A & c^2 \: \mathcal{E}^* & c^2 \: \mathcal{W}_A^* \\ 
0 & c^2 \: \mathcal{E} & c^2 \: \mathcal{P}_B & c^2 \: \mathcal{W}_B \\ 
c^2 \: \mathcal{P}_{AB} & c^2 \: \mathcal{W}_A & c^2 \: \mathcal{W}_B^* & 
1 - c^2 \big( \mathcal{P}_A + \mathcal{P}_B \big) \end{pmatrix} + O(c^4) . 
\label{eqn:PartTransRho} 
\end{equation}  
The eigenvalues of $\rho_{AB}^{T_A}$ are derived by applying again Eq. (\ref{eqn:EigenValuesRhoS}) in Appendix \ref{sec:Eigenvalues} as 
\begin{equation} 
1 + O(c^2) , \quad \dfrac{c^2}{2} \left[ \mathcal{P}_A + \mathcal{P}_B 
\pm \sqrt{\left( \mathcal{P}_A - \mathcal{P}_B \right)^2 + 4 \left| \mathcal{E} \right|^2} \right] + O(c^4) , \quad 
c^4 \left( \mathcal{X} - \left| \mathcal{P}_{AB} \right|^2 \right) + O(c^6) , 
\label{eqn:EigenVPartTransRho} 
\end{equation} 
By noting Eq. (\ref{eqn:PositivePs}) again, the condition for the two detectors to be entangled is then described as 
\begin{subequations}\label{eqn:EntanglementCond}
\begin{align}
&\mathcal{P}_A \, \mathcal{P}_B < \left| \mathcal{E} \right|^2 , \label{eqn:EntanglementCond1}\\
&\qquad  \mathrm{or}  \nonumber \\
& \mathcal{X} < \left| \mathcal{P}_{AB} \right|^2  ,   
\label{eqn:EntanglementCond2} 
\end{align}
\end{subequations} 
where $\mathcal{X}$ is defined in Eq. (\ref{eqn:CalXDef}). 
It is worth mentioning that these conditions coincide exactly with those derived by Reznik for  
the restricted form of monopole coupling \cite{Reznik03}, where $\mathcal{W}_I$ is absent;  
hence those conditions are general enough for the extraction of entanglement from a vacuum as long as one considers a monopole coupling.

We notice that conditions \eqref{eqn:EntanglementCond1} and \eqref{eqn:EntanglementCond2} are not compatible with each other, and hence either of them, but not both, is 
the condition for the two detectors to be entangled. Indeed, by taking into account the positivity (\ref{eqn:PositivityRho}) of 
the reduced density matrix $\rho_{AB}$, 
we find that Eq.~\eqref{eqn:EntanglementCond1} gives 
\begin{equation}
\mathcal{X} \geq \left| \mathcal{E} \right|^2 > \mathcal{P}_A \, \mathcal{P}_B \geq \left| \mathcal{P}_{AB} \right|^2 , 
\label{eqn:EntCondExcld1} 
\end{equation}
which contradicts the second condition \eqref{eqn:EntanglementCond2}, while Eq.~\eqref{eqn:EntanglementCond2} yields 
\begin{equation}
\mathcal{P}_A \, \mathcal{P}_B \geq \left| \mathcal{P}_{AB} \right|^2 > \mathcal{X} \geq \left| \mathcal{E} \right|^2 , 
\label{eqn:EntCondExcld2} 
\end{equation} 
which is incompatible with the first condition \eqref{eqn:EntanglementCond1}. 

In order to discuss the nature of extracted entanglement, we shall compute below several entanglement measures 
for arbitrary switching functions and arbitrary worldlines of the detectors in an arbitrary spacetime.  
In the first two subsections, we will consider the bounds on the distillable entanglement and the entanglement cost. 
In the third subsection, we will turn to the fundamental issue of the Bell-CHSH inequality, and investigate whether Unruh-DeWitt detectors are suitable in the sense of Summers and Werner 
\cite{SummersWerner85,SummersWerner87a,SummersWerner87b,SummersWerner87c} in the general context.
Note that the distillable entanglement and the entanglement cost assume many copies of entangled states, which would be rather unrealistic in our case since one needs to prepare many vacuum states and many pairs of Unruh-DeWitt detectors.
Motivated by this, in the last subsection, we will consider the quantum teleportation that is implemented with the single copy of entangled detectors and compute the teleportation fidelity. 


\subsection{Bounds on distillable entanglement --- Negativity and coherent information} 
\label{subsec:NegativityConcurrence} 

The distillability of singlet states (two-qubit maximally entangled states) is important for many applications, such as quantum key distribution and quantum teleportation. 
While there generally exists an entangled state from which no singlet states can be extracted, i.e., a bound entangled state, it is always possible to distill singlet states from many copies of an arbitrary two-qubit entangled state \cite{HorodeckiHH97}. Hence, in principle, one can distill singlet states from the extracted entangled states also in our case. 
However, beyond such a qualitative discussion, we still need to give a quantitative estimation for the distillability. 
The distillable entanglement $E_D(\rho_{AB})$, defined asymptotically ($n \rightarrow \infty$) as the optimal rate $m/n$ to extract $m$ copies of a singlet state from $n$ copies of a states $\rho_{AB}$ through the local operations and classical communication (LOCC), provides such an operationally motivated measure. 
Unfortunately, the measure is generally known to be difficult to compute even for two-qubit cases. So, here we shall focus on other computable measures that give an upper bound and a lower bound 
of the distillable entanglement. 

The negativity ${\cal N}(\rho_{AB})$ of a density operator $\rho_{AB}$ is defined as minus the sum of the negative eigenvalues of its partial transpose $\rho_{AB}^{T_A}$ \cite{neg}. 
Moreover, the logarithmic negativity $\log_2 \left(2 {\cal N}(\rho_{AB}) +1 \right)$ gives an upper bound of the distillable entanglement \cite{UpBoundDis}:
\begin{equation}\label{eq:bddED} 
E_D(\rho_{AB}) \le \log_2 \left(2 {\cal N}(\rho_{AB}) +1 \right). 
\end{equation} 
For the reduced density matrix derived in Eq. (\ref{eqn:RedDensityMatrix}), we immediately obtain from Eqs. \eqref{eqn:EigenVPartTransRho} and \eqref{eqn:EntanglementCond}, 
\begin{equation}
{\cal N}(\rho_{AB}) = \left\{ \begin{array}{lcc} 
- \dfrac{c^2}{2} \left[ \mathcal{P}_A + \mathcal{P}_B 
- \sqrt{\left( \mathcal{P}_A - \mathcal{P}_B \right)^2 + 4 \left| \mathcal{E} \right|^2} \right] + O(c^4) & \quad \mathrm{for} \  {\rm condition} \  \eqref{eqn:EntanglementCond1}, \\ 
c^4 \left( \left| \mathcal{P}_{AB} \right|^2 -  \mathcal{X}\right) + O(c^6) & \quad \mathrm{for} \ \mathrm{condition} \ \eqref{eqn:EntanglementCond2}. \end{array} \right. 
\label{eqn:Negativity} 
\end{equation} 
Therefore, this gives an upper bound of possible distillation \eqref{eq:bddED}  of singlet states from our extracted entanglement.

Surely, a more interesting estimation is a lower bound of $E_D(\rho_{AB})$, as it will guarantee the amount of extraction of singlet states at lowest. 
It was shown \cite{ref:DW} that the coherent information $I_c(A \ \rangle B)$ (from $B$ to $A$) gives a lower bound of the one-way entanglement capacity $D_{\rightarrow}(\rho_{AB})$, i.e., entanglement distillation with one-way communication: 
\begin{equation}
E_D(\rho_{AB}) \ge D_{\rightarrow}(\rho_{AB}) \ge I_c(A \ \rangle B). 
\end{equation} 
The coherent information $I_c(A \ \rangle B)$ of a state $\rho_{AB}$ is defined by 
\begin{equation}
I_c(A \ \rangle B) \equiv H(\rho_B) - H(\rho_{AB}),
\end{equation}
where $H(\rho) \equiv - \mathrm{Tr} \rho \, \log_2 \rho$ is the von Neumann entropy, and thus $I_c(A \ \rangle B)$ coincides with the negative of the conditional entropy $I_c(A \ \rangle B) = - H(A|B)$. From Eqs. (\ref{eqn:EigenVRho}) and (\ref{eqn:EigenVRhoB}), we obtain, 
for the reduced density matrix (\ref{eqn:RedDensityMatrix}), 
\begin{eqnarray}
I_c(A \ \rangle B) = 2 \, \mathcal{P}_A \, c^2 \, \log_2 | c | + O(c^2) .  
\end{eqnarray} 
Since $|c| < 1$ within the perturbation theory and $\mathcal{P}_A$ is positive \eqref{eqn:PositivePs}, we find that $I_c(A \ \rangle B)$ is negative.  
Therefore, the fact that the excitation probability is non-negative unfortunately prohibits us from obtaining a meaningful lower bound for entanglement distillation. 

\subsection{Entanglement cost --- Entanglement of formation and concurrence}

Another operationally motivated measure of entanglement is the entanglement of cost $E_C(\rho_{AB})$, which is defined as the optimal rate to cost copies of a singlet state in order to obtain the copies of a state $\rho_{AB}$. The related measure is the entanglement of formation $E_F (\rho_{AB})$ \cite{BennettDSW96}:   
\begin{equation}
E_F(\rho_{AB}) \equiv \inf \sum_j p_j E(\phi_j) , 
\end{equation}
where $\inf$ is taken over all possible decompositions into pure states $\cket{\phi_j}$ as $\rho_{AB} = \sum_j p_j \cket{\phi_j}\bra{\phi_j}$, and $E(\phi_j)$ is the entropy of entanglement of a pure state $\cket{\phi_j}$. In particular, $E_C(\rho_{AB})$ 
and  $E_F(\rho_{AB})$ are related as 
\begin{equation} 
E_C(\rho_{AB}) = \lim_{n \to \infty} \frac{E_F(\rho_{AB}^{\otimes n})}{n} .
\end{equation} 
Although both $E_C(\rho_{AB})$ and $E_F(\rho_{AB})$ are generally difficult to compute, it is known \cite{Wootters98} in the case of a two-qubit system that $E_F(\rho_{AB})$ is given by a computable quantity, the concurrence $C(\rho_{AB})$, through the formula  
\begin{equation}
E_F(\rho_{AB}) = h \Bigl( \frac{1 + \sqrt{1-C^2(\rho_{AB})}}{2}\Bigr) , 
\end{equation}
where $h(x) \equiv -x \, \log_2 x - (1-x) \, \log_2 (1-x)$ is the binary entropy.  
The concurrence $C(\rho_{AB})$ for a two-qubit state $\rho_{AB}$ is defined by
\begin{equation}
C(\rho_{AB}) \equiv \max[0, \tilde{\lambda}_1 - \tilde{\lambda}_2 - \tilde{\lambda}_3 - \tilde{\lambda}_4] , 
\label{eqn:ConcurrenceDef} 
\end{equation}
where $\tilde{\lambda}_1$, $\tilde{\lambda}_2$, $\tilde{\lambda}_3$, and $\tilde{\lambda}_4$ are the square roots of the eigenvalues of 
$\rho_{AB} \, \tilde{\rho}_{AB}$ in the descending order and $\tilde{\rho}_{AB}$ is defined as 
\begin{equation}
\tilde{\rho}_{AB} 
\equiv \sigma_y \otimes \sigma_y \, 
\rho^\ast_{AB} \, 
\sigma_y \otimes \sigma_y .  
\label{eqn:TildeRhoDef} 
\end{equation} 
The square roots of the eigenvalues of $\rho_{AB}\tilde{\rho}_{AB}$ for the reduced density matrix (\ref{eqn:RedDensityMatrix}) are calculated from the eigenvalue equation 
(\ref{eqn:EigenValueEqRhoTIlRho}) in Appendix \ref{sec:Eigenvalues} as 
\begin{equation} 
c^2 \left( \sqrt{{\cal P}_A {\cal P}_B} \pm |{\cal P}_{AB}| \right) + O(c^4) , \quad 
c^2 \left( \sqrt{\mathcal{X}} \pm |{\cal E}| \right) + O(c^4) . 
\label{eqnEigenValueRhoTildeRho} 
\end{equation}
As we noticed in Eq. (\ref{eqn:EntCondExcld1}), 
when $\left| \mathcal{E} \right| > \sqrt{\mathcal{P}_A \, \mathcal{P}_B}$, we have 
$\sqrt{\mathcal{X}} > \left| \mathcal{P}_{AB} \right|$, and hence 
$\sqrt{\mathcal{X}} + \left| \mathcal{E} \right| > \sqrt{\mathcal{P}_A \, \mathcal{P}_B} + \left| \mathcal{P}_{AB} \right|$. 
In this case, the maximal eigenvalue $\tilde{\lambda}_1$ is found to be $\tilde{\lambda}_1 = c^2 \left( \sqrt{\mathcal{X}} + |{\cal E}| \right) + O(c^4)$. 
When $\sqrt{\mathcal{X}} < \left| \mathcal{P}_{AB} \right|$, on the other hand, we have 
$\sqrt{\mathcal{P}_A \, \mathcal{P}_B} > \left| \mathcal{E} \right|$ from Eq. (\ref{eqn:EntCondExcld2}), which implies 
$\sqrt{\mathcal{P}_A \, \mathcal{P}_B} + \left| \mathcal{P}_{AB} \right| > \sqrt{\mathcal{X}} + \left| \mathcal{E} \right|$, 
and thus $\tilde{\lambda}_1 = c^2 \left( \sqrt{{\cal P}_A {\cal P}_B} + |{\cal P}_{AB}| \right) + O(c^4)$. 
Therefore, the concurrence $C(\rho_{AB})$ associated with the reduced density matrix 
(\ref{eqn:RedDensityMatrix}) is computed as 
\begin{equation}
C(\rho_{AB}) = \left\{ \begin{array}{lcc} 
2 \, c^2 \left( |{\cal E}| - \sqrt{{\cal P}_A{\cal P}_B} \right) + O(c^4) & \quad \mathrm{for} \ \mathrm{condition} \ \eqref{eqn:EntanglementCond1}, \\ \\ 
2 \, c^2 \left( |{\cal P}_{AB}| - \sqrt{\mathcal{X}} \right) + O(c^4) & \quad \mathrm{for} \ \mathrm{condition} \ \eqref{eqn:EntanglementCond2}. \end{array} \right. 
\label{eqn:Concurrence} 
\end{equation} 
It is interesting to notice that when the second condition \eqref{eqn:EntanglementCond2} for entanglement holds, 
the contribution to the concurrence is at order of $c^2$ while the contribution to the negativity is at order of $c^4$ (see Eq. \eqref{eqn:Negativity}).  
On the other hand, when the first condition \eqref{eqn:EntanglementCond1} is met, 
we see from Eqs. (\ref{eqn:Negativity}) and (\ref{eqn:Concurrence}) that 
in the particular case of $\mathcal{P}_A = \mathcal{P}_B$, which we thus denote as $\mathcal{P}_I$, the concurrence and the negativity are related and given as 
\begin{equation}
C(\rho_{AB}) = 2 \, {\cal N}(\rho_{AB}) = 2 \, c^2 \left( |{\cal E}| - {\cal P}_I \right) + O(c^4) . 
\label{eqn:ConcurrenceNegativity} 
\end{equation} 
The same relation has been derived in Ref. \cite{MartinMartinezST16} for a restricted form of the monopole coupling. 
We will use Eq. (\ref{eqn:ConcurrenceNegativity}) in the last subsection below.

\subsection{Bell-CHSH inequality} 
\label{subsec:Usefulness} 
In this subsection, we shall ask whether the entanglement generated between the Unruh-DeWitt detectors can be explained by classical means, i.e., by the local realism. 
Indeed, Summers and Werner \cite{SummersWerner85,SummersWerner87a,SummersWerner87b,SummersWerner87c} showed the maximal violation of the Bell-CHSH inequality in a vacuum of any quantum field theory, which implies that a vacuum intrinsically includes the non-locality that cannot be explained by any local realistic means. 
If one could observe the non-locality in entanglement generated between detectors separated acausally, it would provide an indirect corroboration of their results.  
Notice, however, that the presence of entanglement does not necessary imply the non-locality, since there is an entangled state the statistics of which can be still reproducible by some local hidden variable models \cite{ref:Werner,ref:EntLR}. 
Therefore, we still need to check the violation of a Bell inequality for the extracted entanglement.
Here, we focus on the so-called Bell-CHSH inequality, which gives not only the necessary condition for the existence of the local hidden variable models \cite{ref:CHSH}, but also the sufficient condition \cite{ref:CHSHfine} in {\it the CHSH setting}, i.e., the physical setting where there are two-valued dichotomic measurements in the bipartite system. 

For a two-qubit state $\rho_{AB}$, letting the ``optimal'' CHSH quantity \cite{ref:HHH} be  
\begin{equation} 
\beta_{\rm CHSH}(\rho_{AB}) 
\equiv \max_{{\bm a},{\bm a'},{\bm b},{\bm b'}} 
\mathrm{Tr} \, \rho_{AB}  
\left( \sigma_{\bm a} \otimes (\sigma_{\bm b}+\sigma_{\bm b'}) + \sigma_{\bm a'} \otimes (\sigma_{\bm b}-\sigma_{\bm b'}) \right) , 
\label{eqn:CHSHBeta} 
\end{equation}
where the maximization in Eq. (\ref{eqn:CHSHBeta}) is taken over all spin observables $\sigma_{\bm a},\sigma_{\bm a'}$ for one qubit and $\sigma_{\bm b},\sigma_{\bm b'}$ for the other, the Bell-CHSH inequality is described as   
\begin{equation}\label{eq:CHSHineq}
\beta_{\rm CHSH}(\rho_{AB})  \le 2. 
\end{equation} 
One can show \cite{ref:HHH} that 
\begin{equation}\label{eq:CHSHM}
\beta_{\rm CHSH}(\rho_{AB})  = 2 
\sqrt{M(\rho_{AB})} ,  
\end{equation}
where $M(\rho_{AB})$ is the sum of the two greatest eigenvalues of 
\begin{equation}\label{eq:U}
U_{\rho_{AB}}  
\equiv 
T^T_{\rho_{AB}} T_{\rho_{AB}} , 
\end{equation}
and $T_{\rho_{AB}}$ is the $3\times 3$ matrix defined by 
\begin{equation}
(T_{\rho_{AB}})_{ij} \equiv \mathrm{Tr} ( \rho_{AB}  \, \sigma_i \otimes \sigma_j). 
\label{eqn:MtrxTDef} 
\end{equation} 
From the eigenvalue equation (\ref{eqn:EigenValueEqU}) in Appendix \ref{sec:Eigenvalues}, the eigenvalues of the matrix $U_{\rho_{AB}}$ for the reduced density matrix (\ref{eqn:RedDensityMatrix}) are derived as   
\begin{equation}\label{eq:eig}
1 - 4 \,c^2 \left( \mathcal{P}_A + \mathcal{P}_B \right) + O(c^4) , \quad 
4 \, c^4 \left( \left| \mathcal{E} \right| \pm \left| \mathcal{P}_{AB} \right| \right)^2 + O(c^6) . 
\end{equation}
Consequently, we have 
\begin{equation} 
M(\rho_{AB}) = 1 - 4 ({\cal P}_A +{\cal P}_B)c^2 + O(c^4),
\end{equation} 
and hence we obtain from Eq. (\ref{eq:CHSHM}), 
\begin{equation}
\beta_{\rm CHSH}(\rho_{AB}) = 2 \left( 1 - 2({\cal P}_A +{\cal P}_B)c^2 \right) + O(c^4) \leq 2 . 
\label{eqn:NonViolationBell} 
\end{equation} 
Therefore, we see that the entanglement generated between the two Unruh-DeWitt detectors does not violate the Bell-CHSH inequality \eqref{eq:CHSHineq}. 
The same conclusion was observed in \cite{ReznikRS05,NambuOhsumi09}. 
We note again that our interaction is more general than those used there; hence the derived result here shows the fact of non-violation of the Bell-CHSH inequality of the extracted entanglement in a more general context. 
Notice, however, that this result is still not sufficient to conclude that the extracted entanglement can be explainable by local realism. Indeed, the optimization in Eq. \eqref{eqn:CHSHBeta} is only through spin observables, while there are more general positive operator valued measure (POVM) measurements. Moreover, as is noticed above, the Bell-CHSH inequality gives the sufficient condition for the local realism only in the restriction to the CHSH setting \footnote{
In \cite{ReznikRS05,NambuOhsumi09}, the authors also showed that there is a hidden non-locality \cite{ref:G} by using a local filtering. However, we should be careful in concluding the nature of non-locality. Similar to the argument of the detection loophole of Bell inequality, it is still not clear that the extracted entanglement cannot be explained by local realism \cite{ref:M}. }. 
In this restricted sense, however, we see again that the non-negativity of the excitation probability (\ref{eqn:PositivePs}) prevents the Bell-CHSH inequality from being violated, and hence Unruh-DeWitt detectors are found not to be suitable detectors in the sense 
of Summers and Werner \cite{SummersWerner85,SummersWerner87a,SummersWerner87b,SummersWerner87c}, also for the general monopole coupling.  

\subsection{Optimal fidelity of teleportation}

In this subsection, we consider the possibility of quantum teleportation only via the single copy of the entangled detectors. 
To do this, we compare the optimal teleportation fidelity with entanglement to that achievable only via classical channels.
The (averaged) teleportation fidelity is given by
\begin{equation} 
F = \int dM(\phi) \sum_k p_k \langle \phi| \rho_k |\phi\rangle , 
\end{equation} 
where the integral is over the quantum state $|\phi \rangle$ to be teleported from the sender with the unitary invariant measure $dM$, and $\rho_k$ is the state teleported to the receiver given the outcome $k$ of the sender's measurement with the probability $p_k$. 
Let $F(\rho_{AB})$ denote the fidelity of teleportation using an entangled state $\rho_{AB}$. 
The Horodecki family showed \cite{ref:HHHTel} the optimal fidelity for a two-qubit system is given as 
\begin{equation}
F_{\max}(\rho_{AB}) 
= \frac{1}{2} \left( 1 + \frac{1}{3} \mathrm{Tr} 
\sqrt{U_{\rho_{AB}}} \right) ,
\label{eqn:FidelityHoro} 
\end{equation}
where $U_{\rho_{AB}}$ is given by Eq. \eqref{eq:U}. 
Here, we are concerned only with the standard teleportation, and then the optimization is over all local unitary operations on the receiver's side while the Bell measurement on the sender's side is fixed. (See \cite{ref:HHHGenTel} for general protocols based on LOCC.)  
On the other hand, one can show \cite{ref:HHHGenTel,ref:ClTel} that the classically achievable fidelity is $F_{cl} =2/3 $. Hence, 
the condition for the supremacy of teleportation using entanglement over classical channels is given by $F_{\max}(\rho_{AB}) > 2/3$, which turns out to be  
\begin{equation}\label{cond:tel} 
\mathrm{Tr} \sqrt{U_{\rho_{AB}}} > 1. 
\end{equation}  
Note that, while the violation of the Bell-CHSH inequality is sufficient to obtain the fidelity of the standard quantum teleportation larger than its 
classical value \cite{ref:HHHTel}, the converse does not hold in general. 
Thus, in spite of the previous result in subsection \ref{subsec:Usefulness}, it is still meaningful to investigate here whether the entanglement generated between the Unruh-DeWitt detectors can be utilized in quantum teleportation. 

By summing the square roots of eigenvalues \eqref{eq:eig}, we find, for the reduced density matrix (\ref{eqn:RedDensityMatrix}), 
\begin{equation} 
\mathrm{Tr} \sqrt{U_{\rho_{AB}}} = \left\{ \begin{array}{ccc} 
1 + 4 \, c^2 \left[ \left| \mathcal{E} \right| - \dfrac{\mathcal{P}_A + \mathcal{P}_B}{2} \right] + O(c^4) & 
\quad \mathrm{for} \quad 
& \left| \mathcal{E} \right| > \left| \mathcal{P}_{AB} \right| \\ 
1 + 4 \, c^2 \left[ \left| \mathcal{P}_{AB} \right| - \dfrac{\mathcal{P}_A + \mathcal{P}_B}{2} \right] + O(c^4) & 
\quad \mathrm{for} \quad 
& \left| \mathcal{E} \right| < \left| \mathcal{P}_{AB} \right| . 
\end{array} \right. 
\end{equation} 
 
If the second condition \eqref{eqn:EntanglementCond2} for entanglement holds, and hence Eq. (\ref{eqn:EntCondExcld2}) is valid, we have $\left| \mathcal{E} \right| < \left| \mathcal{P}_{AB} \right|$. In this case, we have 
\begin{equation}
\mathrm{Tr} \sqrt{U_{\rho_{AB}}} = 1 + 4 \, c^2 \left( \left| \mathcal{P}_{AB} \right| - \dfrac{\mathcal{P}_A + \mathcal{P}_B}{2} \right) + O(c^4). 
\end{equation}
However, Eq.~\eqref{eqn:EntCondExcld2} and $\sqrt{{\cal P}_A {\cal P}_B} \le 
\dfrac{\mathcal{P}_A + \mathcal{P}_B}{2}$ show that condition \eqref{cond:tel} fails. 
Therefore, the entanglement extracted due to the second condition \eqref{eqn:EntanglementCond2} is not useful in quantum teleportation. 

On the other hand, when the first condition \eqref{eqn:EntanglementCond1} is satisfied, and hence 
Eq. (\ref{eqn:EntCondExcld1}) is valid, we have $\left| \mathcal{E} \right| > \left| \mathcal{P}_{AB} \right|$. 
In this case, we have 
\begin{equation}
\mathrm{Tr} \sqrt{U_{\rho_{AB}}} = 1 + 4 \, c^2 \left( \left| \mathcal{E} \right| - \dfrac{\mathcal{P}_A + \mathcal{P}_B}{2} \right) + O(c^4) , 
\label{eqn:NFidelityP} 
\end{equation} 
In particular, when the two detectors are symmetric in the sense $\mathcal{P}_A = \mathcal{P}_B$, which we write as $\mathcal{P}_I$, 
Eq. (\ref{eqn:EntCondExcld1}) shows $\left| \mathcal{E} \right| > \dfrac{\mathcal{P}_A + \mathcal{P}_B}{2} = {\cal P}_I$. Hence the condition \eqref{cond:tel} holds, implying that the extracted entangled state is indeed useful for the standard teleportation. 
Moreover, since the negativity ${\cal N}(\rho_{AB})$ and the concurrence $C(\rho_{AB})$ in this case are given by Eq. \eqref{eqn:ConcurrenceNegativity}, we have  
\begin{equation}\label{eq:opfid}
F_{\max}(\rho_{AB}) 
= \frac{2}{3}  + \frac{2}{3} \, {\cal N}(\rho_{AB}) + O(c^4) = \frac{2}{3}  + \frac{1}{3} \, C(\rho_{AB}) + O(c^4) . 
\end{equation}
Therefore, we have shown that the extracted entanglement with the first condition \eqref{eqn:EntanglementCond1} for the symmetric case is useful actually 
for the standard teleportation only with the use of the single copy of the entangled Unruh-DeWitt detectors. 
Furthermore, we found the interesting characterization \eqref{eq:opfid} of the optimal fidelity expressed in terms of the negativity and the concurrence. 
  
\section{Inertial motions in Minkowski vacuum} 
\label{sec:MinkowskiEn} 

Now that we have seen that entanglement generated between Unruh-DeWitt detectors, if any, is useful in quantum teleportation, our interest is then whether and 
how entanglement is generated between the Unruh-DeWitt detectors. 
In this section, we focus on inertial motions of Unruh-DeWitt detectors in the Minkowski vacuum, where 
the Wightman function $G_W(x, x')$ of a neutral massless scalar field is given by 
\begin{equation} 
G_W(x, x') = \frac{- 1}{( 2 \pi )^2} 
\frac{1}{( t - t' - \ImUnit \, \varepsilon )^2 - | \boldsymbol{x} - \boldsymbol{x}' |^2} ,  
\label{eqn:WightmanGen} 
\end{equation} 
and $t$ and $\boldsymbol{x}$ are Cartesian coordinates, 
and evaluate the entanglement generated between the Unruh-DeWitt detectors. 

We will consider in the first subsection the case where the switching functions $\chi_I(\tau_I)$ in Eqs. (\ref{eqn:RedMatElPI}) and (\ref{eqn:RedMatElE1}) are set as 
$\chi_I(\tau_I) = 1$ identically, while adiabatic switching on and off at infinite past and future is assumed implicitly, as in the textbooks on  
Unruh-DeWitt detectors \cite{DeWitt79,BirrellDavies82}, which we call the implicit adiabatic switching in what follows. 
In the second subsection, we will analyze 
the explicit effects of switching on and off of Unruh-DeWitt detectors, while being concerned only with the case of two comoving 
Unruh-DeWitt detectors.  

\subsection{Implicit adiabatic switching} 
\label{subsec:WithoutSWT} 

In this subsection, we consider the implicit adiabatic switching $\chi_I(\tau_I) \equiv 1$, and thus the integrals 
in Eqs. (\ref{eqn:RedMatElPI}) and (\ref{eqn:RedMatElE1}) reduce to 
\begin{align}  & 
\mathcal{I}_I = 
\int_{- \infty}^{\infty} d \tau'_I \; \int_{- \infty}^{\infty} d \tau_I \; 
e^{\ImUnit \, \Delta E^{(I)} \left( \tau_I -  \tau'_I \right)} \: G_W(\bar{x}'_I , \bar{x}_I) , 
\label{eqn:RedMatElPIChi1} \\ & 
\mathcal{I}_E = 
- \: \ImUnit \, \int_{- \infty}^{\infty} d \tau_B \, \int_{- \infty}^{\infty} d \tau_A \:  
e^{\ImUnit \, \Delta E^{(B)} \tau_B} e^{\ImUnit \, \Delta E^{(A)} \tau_A} \: G_F(\bar{x}_B , \bar{x}_A) .   
\label{eqn:RedMatElE1Chi1}
\end{align} 

For an inertial motion of the detector $I$, the coordinates of the worldline are given by 
\begin{equation}
\bar{t}_I(\tau_I) = \frac{1}{\sqrt{1 - v_{I 0}^2}} \, \tau_I , \qquad 
\bar{\boldsymbol{x}}_I(\tau_I) = \boldsymbol{x}_{I 0} + \frac{\boldsymbol{v}_{I 0}}{\sqrt{1 - v_{I 0}^2}} \, \tau_I  , 
\label{eqn:WLsInertial}
\end{equation} 
where $\boldsymbol{x}_{I 0}$ and $\boldsymbol{v}_{I 0}$ are constants, and $v_{I 0} \equiv | \boldsymbol{v}_{I 0} |$. 
Then the Wightman function $G_W(\bar{x}_I' , \bar{x}_I)$ between the events $\bar{x}_I'$ and $\bar{x}_I$ along the worldline of the detector $I$ reduces to 
\begin{equation}
G_W(\bar{x}_I' , \bar{x}_I) = \frac{- 1}{( 2 \pi )^2} 
\frac{1}{( \tau_I' - \tau_I - \ImUnit \, \varepsilon )^2} ,  
\label{eqn:WightmanIntOn} 
\end{equation}
and thus $\mathcal{I}_I$ is rewritten from Eq. (\ref{eqn:RedMatElPIChi1}) as 
\begin{equation}
\mathcal{I}_I = \frac{- 1}{( 2 \pi )^2} 
\int_{- \infty}^{\infty} d \tau'_I \; \int_{- \infty}^{\infty} d \Delta \tau \; 
e^{\ImUnit \, \Delta E^{(I)} \, \Delta \tau} \: 
\frac{1}{( \Delta \tau + \ImUnit \, \varepsilon )^2} , 
\label{eqn:IIInertialPre} 
\end{equation}
where $\Delta \tau \equiv \tau_I - \tau_I'$. 
By analytically continuing onto the complex $\Delta \tau$ plane and taking as the integral path 
the infinite semicircle in the upper half of the complex $\Delta \tau$ plane, where the integrand in 
Eq. (\ref{eqn:IIInertialPre}) is analytic, we obtain $\mathcal{I}_I = 0$. 
Thus, from Eq. (\ref{eqn:IntFactorsPE}), we have 
\begin{equation} 
\mathcal{P}_I = 0 , 
\label{eqn:NoExcitation} 
\end{equation} 
in this case. Actually, Eq. (\ref{eqn:NoExcitation}) is naturally required to hold for inertial detectors in the Minkowski vacuum if they 
properly probe the vacuum, because 
the Minkowski vacuum should be perceived by inertial observers as containing no excited particles, which means that the excitation probability vanishes. 

We notice also that Eq. (\ref{eqn:NoExcitation}) implies that Eq. (\ref{eqn:EntCondExcld2}) is not satisfied, 
and hence that the second condition \eqref{eqn:EntanglementCond2} for entanglement does not hold.  
(This can be seen also from the facts that $\mathcal{P}_{AB}$ vanishes due to Eqs. (\ref{eqn:KeyToVanishing}) and (\ref{eqn:CalPABDefApp}) below, and that $\mathcal{X}$ is non-negative, as we see from Eq. (\ref{eqn:PositivityRho}). 
For a restricted form of the monopole coupling and the gaussian switching function, the same result has been obtained when the 
two detectors are at rest \cite{MartinMartinezST16}.) 
Therefore, the condition for entanglement in this case reduces to the first condition \eqref{eqn:EntanglementCond1}, which is rewritten as 
\begin{equation} 
\mathcal{E} = \bra{E_{1}^{(B)}} m_B(0) \cket{E_0^{(B)}} \, \bra{E_{1}^{(A)}} m_A(0) \cket{E_0^{(A)}} \, 
\mathcal{I}_E \neq 0 . 
\label{eqn:EntanglementEvalInertial} 
\end{equation} 
When Eq. (\ref{eqn:EntanglementEvalInertial}) holds, since $\mathcal{P}_A = \mathcal{P}_B = 0$, the concurrence and the negativity are related and given by 
Eq. (\ref{eqn:ConcurrenceNegativity}), which is expressed as  
\begin{equation}
C(\rho_{AB}) = 2 \, {\cal N}(\rho_{AB}) = 2 \, c^2 \left| {\cal E} \right| + O(c^4) 
= 2 \, c^2 \left| \bra{E_{1}^{(B)}} m_B(0) \cket{E_0^{(B)}} \right| \, \left| \bra{E_{1}^{(A)}} m_A(0) \cket{E_0^{(A)}} \right| \, 
\left| \mathcal{I}_E \right| + O(c^4) , 
\label{eqn:ConcurrenceComoving} 
\end{equation} 
and then the optimal fidelity $F_{\max}(\rho_{AB})$ of quantum teleportation is computed by Eq. (\ref{eq:opfid}), with 
$C(\rho_{AB})$ and ${\cal N}(\rho_{AB})$ being proportional to $\left| \mathcal{I}_E \right|$ by Eq. (\ref{eqn:ConcurrenceComoving}). 
Therefore, we see that whenever $\mathcal{I}_E$ is non-vanishing, the entanglement is generated between the Unruh-DeWitt detectors,  
and it is utilized in quantum teleportation.  

In the remainder of this section, we choose the Cartesian coordinate system where Alice is at rest at the origin, 
and therefore 
\begin{equation} 
\bar{t}_A(\tau_A) = \tau_A , \quad \bar{\boldsymbol{x}}_{A}(\tau_A) = 0 . 
\label{eqn:WLAliceAtRest} 
\end{equation} 
In this case, the Wightman function $G_W(\bar{x}_B , \bar{x}_A)$ 
is written as 
\begin{equation}
G_W(\bar{x}_B , \bar{x}_A) = \frac{- 1}{( 2 \pi )^2} 
\frac{1}{\left( \tau_A - \bar{t}_B(\tau_B) + \ImUnit \, \varepsilon \right)^2 - | \boldsymbol{x}_B(\tau_B) |^2} . 
\end{equation}
Since the poles are located in the lower half of the complex $\tau_A$ plane, by taking the infinite semicircle in the upper half as above, we obtain  
\begin{equation}
\int_{- \infty}^{\infty} d \tau_A \:  
e^{\ImUnit \, \Delta E^{(A)} \tau_A} \: G_W(\bar{x}_B , \bar{x}_A) = 0 . 
\label{eqn:KeyToVanishing} 
\end{equation}
By noticing further that the Feynman propagator $G_F(x , x')$ is decomposed 
(see Appendix \ref{sec:ExplicitMatrixElements}), as 
\begin{equation}
G_F(x , x') = - \, \ImUnit \, G_W(x , x') + G_R(x' , x) , 
\label{eqn:FeynmanDecomp} 
\end{equation} 
we see that $\mathcal{I}_E$ given by Eq. (\ref{eqn:RedMatElE1Chi1}) is computed by 
\begin{equation}
\mathcal{I}_E = 
- \: \ImUnit \, \int_{- \infty}^{\infty} d \tau_B \, e^{\ImUnit \, \Delta E^{(B)} \tau_B} \int_{- \infty}^{\infty} d \tau_A \:  
e^{\ImUnit \, \Delta E^{(A)} \tau_A} \: G_R(\bar{x}_A , \bar{x}_B) , 
\label{eqn:CalIEInertial} 
\end{equation}
where the retarded Green function $G_R(x, x')$ of a neural massless scalar field in 
the Minkowski spacetime is given as 
\begin{equation}
G_R(x, x') = - \, \frac{1}{2 \pi}  \, \Theta(t - t') \, \delta((t - t')^2 - | \boldsymbol{x} - \boldsymbol{x}' |^2) 
= - \frac{1}{4 \pi} \frac{1}{| \boldsymbol{x} - \boldsymbol{x}' |}  \delta(t - t' - | \boldsymbol{x} - \boldsymbol{x}' |)  . 
\label{eqn:RetardedGreenFnc} 
\end{equation}
Therefore, contrary to the literature on entanglement harvesting, e.g., Ref. \cite{Reznik03}, where the contribution to entanglement results only from 
the Wightman function, entanglement extraction is possible only due to the causal propagation of the quantum field described by the retarded Green function in the case of the implicit adiabatic switching, while the quantum correlations due to vacuum fluctuation i.e., virtual processes of the quantum field described by the Wightman function, do not contribute to entanglement extraction.

\subsubsection{Comoving inertial motion} 

Here we consider the case where Alice and Bob are comoving. 
By choosing the coordinate system (\ref{eqn:WLAliceAtRest}), 
Bob also is at rest, and then the worldline of Bob is described as  
\begin{equation} 
\bar{t}_B(\tau_B) = \tau_B , \quad \bar{\boldsymbol{x}}_B(\tau_B) = \boldsymbol{x}_{0} , 
\label{eqn:WLBobAtRest} 
\end{equation} 
with $\boldsymbol{x}_{0} \neq  0$ being constant, and we denote as $L = \left| \boldsymbol{x}_{0} \right|$. 
By substituting Eqs. (\ref{eqn:WLAliceAtRest}) and (\ref{eqn:WLBobAtRest}) into Eqs. (\ref{eqn:CalIEInertial}) and (\ref{eqn:RetardedGreenFnc}), 
we obtain  
\begin{equation}
\mathcal{I}_E = - \frac{1}{2} \: \frac{e^{\ImUnit \, \Delta E^{(A)} L}}{L} \: 
\delta(\Delta E^{(A)} + \Delta E^{(B)}) = 0 , 
\label{eqn:EntglBothAtRest} 
\end{equation}
where one notes $\Delta E^{(I)} > 0$. From Eq. (\ref{eqn:EntanglementEvalInertial}), we thus find that 
entanglement is not extracted in this case. As the appearance of the delta function 
in Eq. (\ref{eqn:EntglBothAtRest}) indicates, when considered as the distribution of the variable $\Delta E^{(A)} + \Delta E^{(B)}$, 
the reason why entanglement is not extracted in this case 
is understood from energy conservation as the consequence of an infinite amount of interaction time (transition time) in 
the uncertainty relation between time and energy. Transitions that violate energy conservation are not allowed after infinitely long interaction, and then the state of the two detectors is forced to return back to the ground state, which is not entangled. 

\subsubsection{Relative inertial motion} 

We next consider the case where Bob is in a relative inertial motion with respect to Alice, 
with a non-vanishing constant relative three-velocity $\boldsymbol{v}_0$, which we assume not parallel or anti-parallel to 
$\boldsymbol{x}_0 \neq 0$ so 
that the worldlines of Alice and Bob do not intersect. 
In this case, the worldline of Bob is given by 
\begin{equation}
\bar{t}_B(\tau_B) = \frac{1}{\sqrt{1 - v^2_0}} \, \tau_B , \qquad 
\bar{\boldsymbol{x}}_B(\tau_B) = \boldsymbol{x}_{0} + \frac{\boldsymbol{v}_0}{\sqrt{1 - v^2_0}} \, \tau_B ,  
\label{eqn:WLBobInMotion}  
\end{equation} 
where $v_0 \equiv \left| \boldsymbol{v}_0 \right|$. 
By substituting Eqs. (\ref{eqn:WLAliceAtRest}) and (\ref{eqn:WLBobInMotion}) into Eqs. (\ref{eqn:CalIEInertial}) and (\ref{eqn:RetardedGreenFnc}), we obtain 
\begin{equation}
\mathcal{I}_E = \frac{\ImUnit}{4 \pi} \int^{\infty}_{- \infty} d \tau_B \, \frac{1}{\left| \bar{\boldsymbol{x}}_B(\tau_B) \right|} 
\, e^{\ImUnit \, \epsilon \, \tau_B} \, e^{\ImUnit \, \Delta E^{(A)} \left| \bar{\boldsymbol{x}}_B(\tau_B) \right|} ,  
\label{eqn:CalIEBobInMotionMid} 
\end{equation} 
where 
\begin{equation}
\epsilon \equiv \Delta E^{(B)} + \frac{1}{\sqrt{1 - v^2_0}} \, \Delta E^{(A)} 
\end{equation}
is the sum of the excitation energies of the two detectors measured in Bob's Lorentz frame. 
By changing the integration variable into $\xi$ defined as 
\begin{equation}
\xi \equiv \tau_B + \frac{\sqrt{1 - v^2_0}}{v_0} | \boldsymbol{x}_0 | \cos \theta ,
\end{equation} 
Eq. (\ref{eqn:CalIEBobInMotionMid}) is written as  
\begin{equation} 
\mathcal{I}_E 
= \frac{\ImUnit}{4 \pi} \frac{\sqrt{1 - v^2_0}}{v_0} \: 
\exp \left[ -  \, \ImUnit \, \frac{\sqrt{1 - v^2_0}}{v_0} \, \epsilon \, | \boldsymbol{x}_0 | \cos \theta \right] 
\int_{- \infty}^{\infty} d \xi \: \frac{1}{\sqrt{ \xi^2 + \ell^2}} 
\, e^{\ImUnit \, \epsilon \, \xi} \, 
e^{\ImUnit \, p \, \sqrt{ \xi^2 + \ell^2}} , 
\label{eqn:CalIEBobInMotionInt} 
\end{equation} 
where $p$ and $\ell$ are defined by  
\begin{equation}
p \equiv \frac{v_0}{\sqrt{1 - v^2_0}} \Delta E^{(A)} , \quad 
\ell \equiv \frac{\sqrt{1 - v^2_0}}{v_0} | \boldsymbol{x}_0 | \sin \theta . 
\end{equation} 
The integral in Eq. (\ref{eqn:CalIEBobInMotionInt}) is 
the zeroth modified Bessel function of the second kind $K_0(\eta)$ of the variable $\eta$ defined as 
\begin{equation} 
\eta \equiv \ell \sqrt{\epsilon^2 - p^2} 
= \frac{\sqrt{1 - v^2_0}}{v_0} \:  | \boldsymbol{x}_0 | \sin \theta \: \sqrt{{\Delta E^{(A)}}^2 
+ \frac{2}{\sqrt{1 - v^2_0}} \Delta E^{(A)} \Delta E^{(B)} + {\Delta E^{(B)}}^2} , 
\label{eqn:EtaDef} 
\end{equation} 
and $\theta \, ( \neq 0 , \pi )$ is the angle between $\boldsymbol{x}_0$ and $\boldsymbol{v}_0$. 
From Eq. (\ref{eqn:WLBobInMotion}), we see that $| \boldsymbol{x}_0 | \sin \theta$  is 
the closest approach between Alice and Bob. Therefore, $\eta$ given by Eq. (\ref{eqn:EtaDef}) is 
found to be invariantly defined under Lorentz transformations and affine reparametrizations. 
We thus obtain 
\begin{equation} 
\left| \mathcal{I}_E \right| =  \frac{1}{2 \pi} \frac{\sqrt{1 - v^2_0}}{v_0} \, K_0(\eta) . 
\end{equation} 
We note that if $\left| \bar{\boldsymbol{x}}_B(\tau_B) \right|$ were constant in Eq. (\ref{eqn:CalIEBobInMotionMid}), 
$\mathcal{I}_E$ would be proportional to the delta function of $\epsilon$ as in the comoving case. 
Thus, the variable distance between Alice and Bob is responsible for the appearance of the modified Bessel function, 
instead of the delta function. 

\begin{figure} 
\includegraphics[scale=0.4]{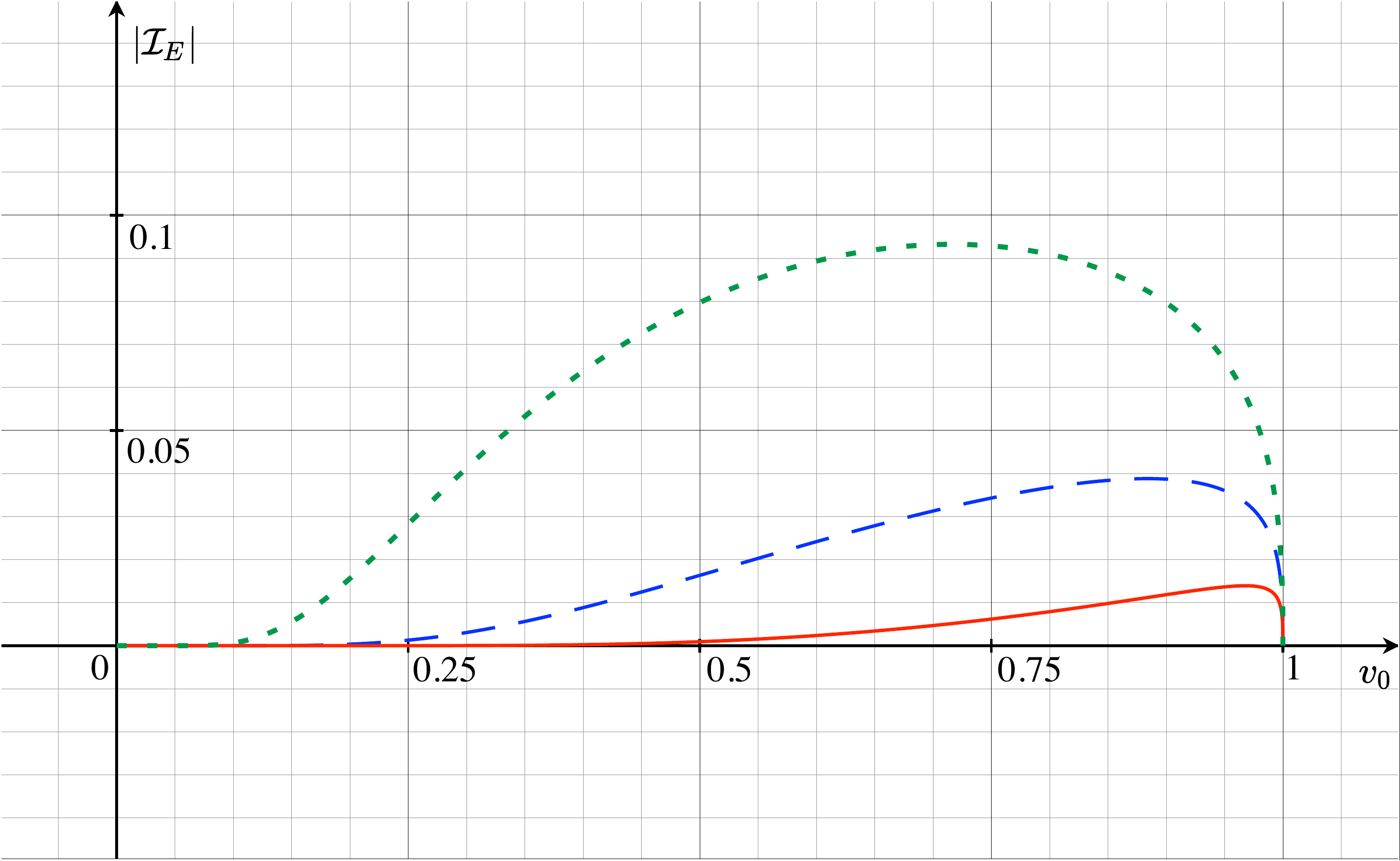} 
\caption{The behavior of $\left| \mathcal{I}_E \right|$, where the horizontal axis is $v_0$. The red solid line denotes $a = 2$, 
the blue dashed line $a = 1$, and the green dotted line $a = 0.5$.} 
\label{fig:EntRelativeMotion}
\end{figure}

From the asymptotic behavior of $K_0(\eta)$ for $\eta \rightarrow \infty$, we immediately see that $\left| \mathcal{I}_E \right|$ vanishes and thus entanglement is not extracted in the comoving case $v_0 = 0$, as we have found above.  
On the other hand, since the leading behavior of $K_0(\eta)$ near $\eta = 0$ is given as $K_0(\eta) \sim - \ln \eta$, 
$\left| \mathcal{I}_E \right|$ drops to zero also in the limit of the speed of light $v_0 \rightarrow 1$. 
This will be naturally understood because the mutual causal contact between Alice and Bob diminishes in this limit. 
However, $\left| \mathcal{I}_E \right|$ does not vanish for $0 < v_0 < 1$. 
To demonstrate this simply, we consider the case of $\Delta E^{(A)} =  \Delta E^{(B)}$, 
which we set as $\Delta E$. Then, $\eta$ is written as 
\begin{equation}
\eta = \frac{a}{v_0} \sqrt{1 - v_0^2 + \sqrt{1 - v_0^2}} , 
\end{equation} 
where $a \equiv \sqrt{2} \: | \boldsymbol{x}_0 | \sin \theta \, \Delta E$ is the ratio of the closest approach to the de Broglie wavelength of a scalar particle with the resonance energy $\Delta E$. 
The behavior of $\left| \mathcal{I}_E \right|$ in this case is shown in Fig. \ref{fig:EntRelativeMotion}.  
As long as the relative velocity is non-relativistic $v_0 \ll 1$, we see from Fig. \ref{fig:EntRelativeMotion} 
that $\left| \mathcal{I}_E \right|$ remains quite small. As the relative velocity becomes relativistic, 
$\left| \mathcal{I}_E \right|$ starts to grow in an accelerated manner, which is thus certainly ascribed to the special relativistic effect. 
Although $\left| \mathcal{I}_E \right|$ decreases quickly as one further increases $v_0$ to the limit of $v_0 \rightarrow 1$, 
we see that the entanglement is extracted and enhanced by the special relativistic effect, unless the relative velocity is 
ultra-relativistic. 
From Eqs. (\ref{eq:opfid}) and (\ref{eqn:ConcurrenceComoving}), we thus find that 
the special relativistic effect enables quantum teleportation. 

\subsection{Switching effects} 
\label{subsec:SwitchEff} 

\begin{figure} 
\includegraphics[scale=0.4,keepaspectratio]{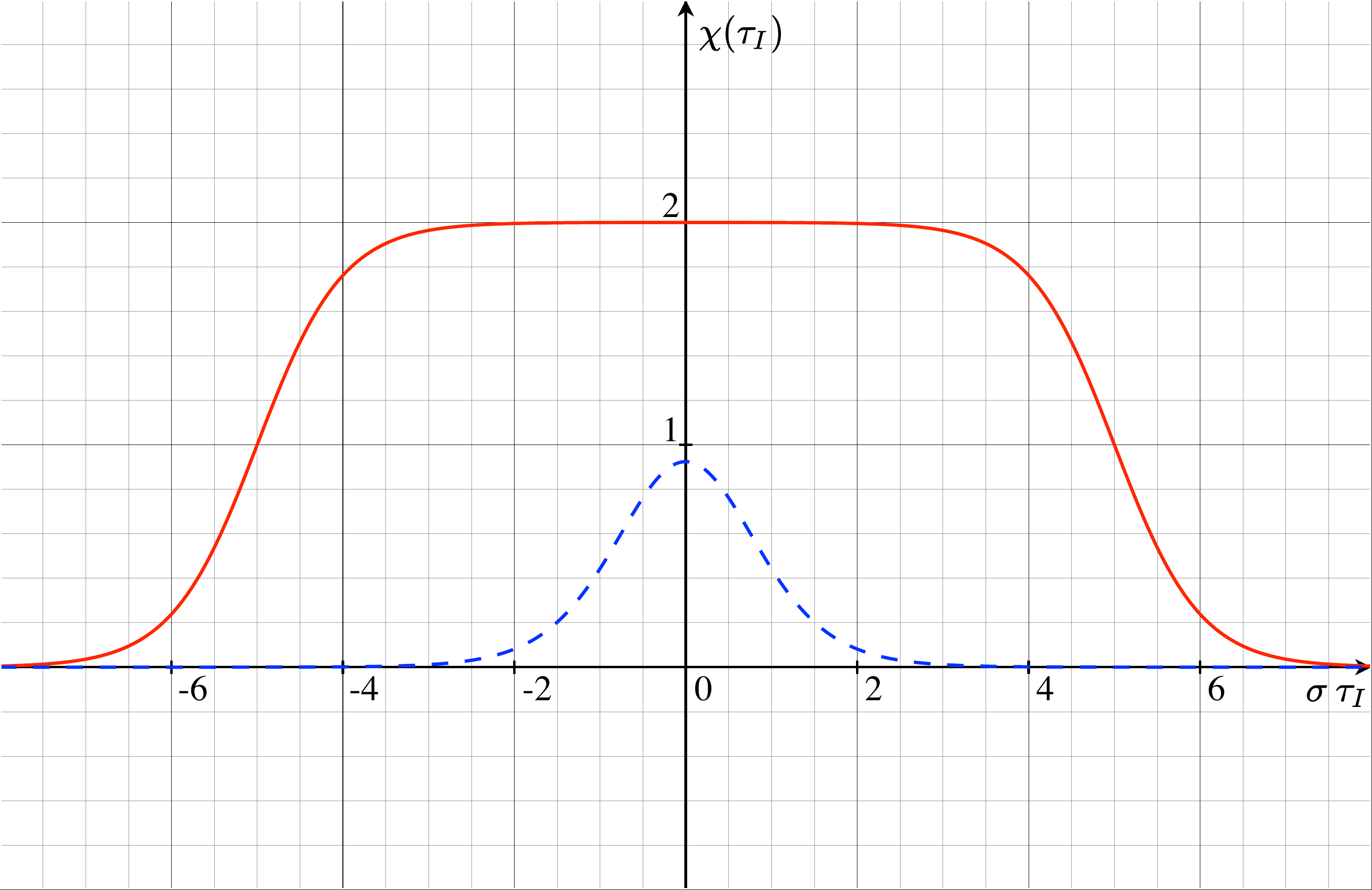} 
\caption{The behavior of the switching function $\chi(\tau_I)$. The horizontal axis denotes $\tau_I$ normalized by $\sigma$ and the vertical axis shows $\chi(\tau_I)$. The blue dashed line shows the behavior of $\chi(\tau_I)$ for $\sigma \, T = 0.5$, and the red solid line for $\sigma \, T = 5$.} 
\label{fig:SwitchingFunc} 
\end{figure}

Now we analyze the effects on entanglement 
caused by switching on and off of Unruh-DeWitt detectors, by focusing on the case where Alice and Bob are comoving. 
For simplicity, we assume that the two Unruh-DeWitt detectors have identical structure and hence set $E_n^{(A)} = E_n^{(B)} \equiv E_n$, $\Delta E \equiv E_1 - E_0$, 
$\cket{E_n} \equiv \cket{E_n^{(I)}}$,  
and $m_A(\tau) = m_B(\tau) \equiv m(\tau)$. 
Furthermore, in order to take into account the effects of switching, we also prescribe the form of the switching function $\chi_I(\tau_I)$ as 
\begin{equation}
\chi_I(\tau_I) = \chi(\tau_I) 
\equiv \tanh \left[ \sigma \left( \tau_I +T \right) \right] 
- \tanh \left[ \sigma \left( \tau_I - T \right) \right] ,   
\label{eqn:SwitchingFuncTanh} 
\end{equation} 
where the positive constants $1 / \sigma$ and $2 T$ denote the timescales of the switching on and off, and the duration that 
$\chi(\tau_I)$ stays above a half of its maximal value, which we call the effective interaction time, 
respectively. Thus, the timescale of the total interaction time is given by 
$2 ( T + 1 / \sigma ) = ( 2 / \sigma ) \left( \sigma T + 1 \right)$. 
In Fig. \ref{fig:SwitchingFunc}, we show two typical cases of the behavior of $\chi(\tau_I)$. 

In this case, we have $\mathcal{P}_A = \mathcal{P}_B$, and hence the concurrence $C(\rho_{AB})$ 
and the negativity ${\cal N}(\rho_{AB})$ when the first condition \eqref{eqn:EntanglementCond1} for entanglement is valid 
are now given from Eq. (\ref{eqn:ConcurrenceNegativity}) as 
\begin{equation}
C(\rho_{AB}) = 2 \, {\cal N}(\rho_{AB}) = 2 \, c^2 \left( |{\cal E}| - {\cal P}_I \right) + O(c^4) 
= 
2 \, c^2 \, \left| \bra{E_{1}} \, m(0) \, \cket{E_0} \right|^2 \left( \left| \mathcal{I}_E \right| - \mathcal{I}_I \right) 
+ O(c^4) , 
\label{eqn:ConcNegSwitchEff}
\end{equation} 
and the fidelity of quantum teleportation is given by Eq. (\ref{eq:opfid}), with $C(\rho_{AB})$ and ${\cal N}(\rho_{AB})$ taking the form of Eq. (\ref{eqn:ConcNegSwitchEff}).  
Since quantum teleportation is not possible in the case of the second condition \eqref{eqn:EntanglementCond2} for entanglement, 
we focus in this subsection on the first condition \eqref{eqn:EntanglementCond1}, which is rewritten in the present case as 
$\left| \mathcal{I}_E \right| - \mathcal{I}_I > 0$, and thus compute and analyze  
the behavior of $\left| \mathcal{I}_E \right| - \mathcal{I}_I$. 

\subsubsection{Computation} 

We first evaluate $\mathcal{I}_I$ 
by employing the expansion of $\tanh z$ as  
\begin{equation} 
\tanh z 
= \sum_{k = 1}^{\infty} \left[ \frac{1}{z + i \left( k - \frac{1}{2} \right) \pi} 
+ \frac{1}{z - i \left( k - \frac{1}{2} \right) \pi} \right]  . 
\label{eqn:TanHFracDecomp} 
\end{equation} 
By substituting Eqs. (\ref{eqn:WightmanGen}), (\ref{eqn:WLAliceAtRest}), 
(\ref{eqn:WLBobAtRest}), (\ref{eqn:SwitchingFuncTanh}), and (\ref{eqn:TanHFracDecomp}) 
into Eq. (\ref{eqn:RedMatElPI}), we perform the first integration by considering the infinite semicircle 
in the upper half of the complex $\tau_I$ plane, within which only the poles of the switching function $\chi(\tau_I)$ 
located at 
\begin{equation} 
\tau_I = \mp T + \ImUnit \left( k - \frac{1}{2} \right) \frac{\pi}{\sigma} 
\end{equation} 
contribute to the integral. 
The second integration in Eq. (\ref{eqn:RedMatElPI}) is performed by taking as the integration path the infinite semicircle 
in the lower half of the complex $\tau_I'$ plane. Again, only the poles of the switching function $\chi(\tau_I')$ 
contribute, which are located at 
\begin{equation} 
\tau_I' = \mp T - \ImUnit \left( k' - \frac{1}{2} \right) \frac{\pi}{\sigma} , 
\end{equation} 
where $k'$ is the summation index that appears in the expansion of $\chi(\tau_I')$ 
as Eq. (\ref{eqn:TanHFracDecomp}). 
Relabeling the summation indices as  $\ell = k + k' -1$ and $m = k - k'$, and performing the summation over $m$, 
we obtain 
\begin{align}
\mathcal{I}_I & = \frac{1}{\pi^2} \left[ - 2 \ln \left( 1 - e^{- \, \frac{\Delta E}{\sigma} \, \pi} \right) 
- e^{2 \, \ImUnit \, T \, \Delta E} \, \Phi\left(e^{- \, \frac{\Delta E}{\sigma} \pi} , - 2 \, \ImUnit \, T \, \frac{\sigma}{\pi} , 1\right) 
- e^{- \, 2 \,\ImUnit \, T \, \Delta E} \, \Phi\left(e^{- \, \frac{\Delta E}{\sigma} \pi} , 2 \, \ImUnit \, T \, \frac{\sigma}{\pi} , 1\right) 
\right. \notag \\ & \left. 
+ \: 2 \,\ImUnit \, T \, \frac{\sigma}{\pi} \left\{ e^{- \, 2 \,\ImUnit \, T \, \Delta E} \, 
\Phi\left(e^{- \, \frac{\Delta E}{\sigma} \pi} , 2 \, \ImUnit \, T \, \frac{\sigma}{\pi} , 2\right)  
- e^{2 \, \ImUnit \, T \,\Delta E} \, \Phi\left(e^{- \, \frac{\Delta E}{\sigma} \pi} , - 2 \, \ImUnit \, T \, \frac{\sigma}{\pi} , 2\right) 
\right\} \right] , 
\label{eqn:CalIIFinApp} 
\end{align}
where $\Phi(z, a, s)$ is the 
Hurwitz-Lerch zeta function defined as 
\begin{equation}
\Phi(z, a , s) \equiv \sum_{n = 0}^{\infty} \frac{z^n}{( a + n )^s} . 
\label{eqn:HurwitzLerchZetaFunc} 
\end{equation} 

By decomposing the Feynman propagator $G_F(x , x')$ in Eq. (\ref{eqn:RedMatElE1}) into 
the Wightman function $G_W(x, x')$ and 
the retarded Green function $G_R(x' , x)$ as Eq. (\ref{eqn:FeynmanDecomp}), we then compute $\mathcal{I}_E$. 
The contribution from the Wightman function $G_W(x , x')$ to $\mathcal{I}_E$ is calculated similarly to the case 
of $\mathcal{I}_I$ above, except that the integral path in the second integration is chosen to be the infinite semicircle in the upper half 
of the complex $\tau_B$ plane, which encircles the five series of poles located at 
\begin{align} & 
\tau_B = \mp \, T + \ImUnit \left( k' - \frac{1}{2} \right) \frac{\pi}{\sigma} , 
\quad
\tau_B =  - L - T + \ImUnit \left\{ \left( k - \frac{1}{2} \right) \frac{\pi}{\sigma} + \varepsilon \right\} , 
\quad
\tau_B =  L - T + \ImUnit \left\{ \left( k - \frac{1}{2} \right) \frac{\pi}{\sigma} +\varepsilon \right\} , 
\notag \\ & 
\tau_B = - L + T + \ImUnit \left\{ \left( k - \frac{1}{2} \right) \frac{\pi}{\sigma} + \varepsilon \right\} , 
\quad 
\tau_B = L + T + \ImUnit \left\{ \left( k - \frac{1}{2} \right) \frac{\pi}{\sigma} +\varepsilon \right\} , 
\label{eqn:PolesCalIKInt}  
\end{align} 
where $L = | \boldsymbol{x}_{0} |$ as above, and $k$ and $k'$ are the summation indices in the expansion of  
the switching function $\chi(\tau_A)$ and $\chi(\tau_B)$, respectively. 
We implement this integration by assuming that these poles do not 
coincide with each other. This requires $T \neq L$ and $2 T \neq L$, but $\mathcal{I}_E$ for these cases is 
determined by continuity. 
We note that the expansion 
\begin{equation}
\coth z = \frac{1}{z} + \sum_{\ell = 1}^{\infty} \left[ \frac{1}{z + i \, \ell \pi} + \frac{1}{z - i \, \ell \pi} \right] 
= \sum_{\ell = - \infty}^{\infty} \frac{z}{z^2 + \ell^2 \pi^2} , 
\label{eqn:CotHFracDecomp} 
\end{equation} 
is helpful in order to simplify the expression. 

On the other hand, the contribution from the retarded Green function $G_R(x' , x)$ to $\mathcal{I}_E$ is 
evaluated by substituting Eqs. (\ref{eqn:WLAliceAtRest}), (\ref{eqn:RetardedGreenFnc}), 
(\ref{eqn:WLBobAtRest}), (\ref{eqn:SwitchingFuncTanh}), and (\ref{eqn:TanHFracDecomp}) 
into Eq. (\ref{eqn:RedMatElE1}), and by using 
\begin{align} & 
\int^{\infty}_{- \infty} d \tau \, 
e^{\ImUnit \, E \, \tau} 
\tanh \left[ \sigma \left( \tau +T_1 \right) \right] \, \tanh \left[ \sigma \left( \tau + T_2 \right) \right] 
\notag \\ & 
= - \frac{2 \, \pi}{\sigma} \frac{1}{\sinh \left[ \frac{\pi \, E}{2 \, \sigma} \right]} \coth \left[ \sigma \left( T_2 - T_1 \right) \right] \, 
\sin \left[ E \, \frac{T_2 - T_1}{2} \right] \, e^{ - \ImUnit \, E \, \frac{T_1 + T_2}{2}} - 2 \, \pi \, \ImUnit \, \delta(E) ,  
\label{eqn:IREvalDerAppFin} 
\end{align} 
which is derived by considering the rectangle with the infinite width (along the real axis) and the height $\pi / \sigma$ (along the imaginary axis) in the upper 
half of the complex $\tau$ plane. By adding these two parts, we finally obtain 
\begin{align} 
\mathcal{I}_E & = \frac{1}{2 \, \sinh \left( \frac{\pi}{\sigma} \Delta E \right)} \frac{1}{\sigma L} 
\left( - \frac{2 L}{\sigma (4 \, T^2 - L^2 )} - 2 \cos \left( 2 \, \Delta E \, T \right) \left[ \frac{1}{\sigma \, L} 
+ \frac{\ImUnit}{\pi} \left\{ \Phi\left(e^{- \frac{\pi}{\sigma} \Delta E} , - \ImUnit \frac{\sigma}{\pi} L , 1 \right) 
- \mathrm{c.c.} 
\right\} \right] 
\right. \notag \\ & \left. 
+ e^{\ImUnit \, \Delta E \, L} \left[ \coth [ \sigma ( 2 \, T + L ) ] e^{2 \, \ImUnit \, \Delta E \, T} 
- \coth [ \sigma ( 2 \, T - L ) ] e^{- \, 2 \, \ImUnit \, \Delta E \, T} - 2 \cos ( 2 \, \Delta E \, T ) \coth( \sigma \, L ) \right] 
\right. \notag \\ & \left. 
- \frac{\ImUnit}{\pi} \left[ \left\{ \Phi\left(e^{- \frac{\pi}{\sigma} \Delta E} , \ImUnit \frac{\sigma}{\pi} ( 2 T + L ) , 1 \right) 
- \Phi\left(e^{- \frac{\pi}{\sigma} \Delta E} , \ImUnit \frac{\sigma}{\pi} ( 2 T - L ) , 1 \right) \right\} 
- \mathrm{c.c.} 
 \right] \right) . 
\label{eqn:CalIEWithSwitch} 
\end{align} 

\subsubsection{Behavior of entanglement} 

When the switching of the detectors is executed quickly enough, it will disturb the quantum state of the scalar field 
and excite the detectors. Since the timescale of the switching is given by $1 / \sigma$, we may apply in this case the approximation 
$\Delta E / \sigma \ll 1$. 
When we keep $\Delta E \, T$ fixed, the Hurwitz--Lerch zeta functions are shown to be bounded, and hence 
Eq. (\ref{eqn:CalIIFinApp}) is approximated as 
\begin{equation} 
\mathcal{I}_I \simeq - \frac{2}{\pi^2} \ln \left( \frac{\pi}{\sigma} \, \Delta E \right) . 
\end{equation} 
Thus, $\mathcal{I}_I$ logarithmically diverges in the limit of $\Delta E / \sigma \rightarrow 0$, 
as in Ref. \cite{LoukoSatz08}.  
The two factors in front of the outermost 
round bracket in Eq. (\ref{eqn:CalIEWithSwitch}) are approximated as $ 1  / ( 2 \, \Delta E \, L )$, and 
$\sigma \, T = \Delta E \, T / ( \Delta E / \sigma ) \rightarrow \infty$ in this limit. Therefore, unless we set the distance between the two detectors to be vanishingly small, 
$\mathcal{I}_E$ remains finite in this limit.  
Although entanglement will be naturally generated between detectors put so close, we see   
that the first condition \eqref{eqn:EntanglementCond1} for entanglement extraction 
is not satisfied under physically plausible circumstances of the finite distance   
for the sudden switching limit $\Delta E / \sigma \rightarrow 0$. This is  consistent with 
Ref. \cite{Pozas-KerstjensMartinMartinez15}, where entanglement is shown not to be extracted 
in the sudden switching limit, while for a different form of the switching function.  

On the other hand, when the switching is performed adiabatically compared to the excitation energy $\Delta E$, we employ the approximation $\Delta E / \sigma \gg 1$. 
In this case, Eq. (\ref{eqn:CalIIFinApp}) is approximated as 
\begin{equation}  
\mathcal{I}_I \simeq e^{- \, \frac{\Delta E}{\sigma} \, \pi} \left[ \frac{2}{\pi^2} 
- 2 \frac{\pi^2 - 4 \, \sigma^2 \, T^2}{( \pi^2 + 4 \, \sigma^2 \, T^2 )^2} \, 
\cos \left( 2 \, \Delta E \, T \right) 
+ \frac{8 \, \pi \, \sigma \, T}{( \pi^2 + 4 \, \sigma^2 \, T^2 )^2} \, 
\sin \left( 2 \, \Delta E \, T \right) \right] , 
\label{eqn:CalIAdiabaitc} 
\end{equation} 
while the approximate form of Eq. (\ref{eqn:CalIEWithSwitch}) is given by 
\begin{align} 
\mathcal{I}_E & \simeq e^{- \, \frac{\Delta E}{\sigma} \, \pi} \, \frac{1}{\sigma \, L} \Big( \frac{2 \, L}{\sigma ( 4 \, T^2 - L^2 )} 
+ \frac{2}{\sigma \, L} \cos ( 2 \, \Delta E \, T ) 
\notag \\ & 
+ \: e^{\ImUnit \, \Delta E \, L} \left\{ \coth [ \sigma ( 2 \, T + L ) ] \, e^{2 \, \ImUnit \, \Delta E \, T} 
- \coth [ \sigma ( 2 \, T - L ) ] \, e^{- \, 2 \, \ImUnit \, \Delta E \, T}  
- 2 \, \cos ( 2 \, \Delta E \, T ) \, \coth ( \sigma \, L ) \right\} \Big) . 
\label{eqn:CalEAdiabatic} 
\end{align} 
In particular, in the limit of $\sigma \, T \rightarrow 0$, by taking the limit $\Delta E / \sigma \rightarrow \infty$ while keeping $\Delta E \, T$ small, Eqs. (\ref{eqn:CalIAdiabaitc}) and (\ref{eqn:CalEAdiabatic}) yield 
\begin{equation}
\left| \mathcal{I}_E \right| - \mathcal{I}_I = \frac{4}{\pi^2} \, e^{- \, \frac{\Delta E}{\sigma} \, \pi} \, \left[ \left( \frac{\pi}{\sigma \, L} \right)^2 - 1 \right] \, \sin^2 \left( \Delta E \, T \right) . 
\label{eqn:EntAdiabiticLimit} 
\end{equation}
We thus see that the entanglement generated between the detectors falls off as $L^{-2}$, which is understood from the behavior of the Wightman function $G_W(x,x')$ in Eq. (\ref{eqn:WightmanGen}) for the massless scalar field, whose Compton wavelength is infinite. Since the timescale of the total interaction time $( 2 / \sigma ) \left( \sigma T + 1 \right)$ (see above) in this case is approximated as $2 / \sigma$, and 
$\left| \mathcal{I}_E \right| - \mathcal{I}_I$ is positive if $L < \pi / \sigma = \left( 2 / \sigma \right) ( \pi / 2 )$, 
we also confirm that the entanglement is generated between the detectors even if they are separated acausally $L > 2 / \sigma$, i.e., even when one of them is put outside the other's lightcone within the total interaction time. Although the switching function $\chi(\tau_I)$ in this paper has 
an exponential tail, our analytical treatment complements the numerical investigation by Reznik \cite{Reznik03}, where the switching function is non-vanishing only for a strictly finite period and hence the detectors are separated acausally in the rigorous sense. 
We note from Eqs. (\ref{eq:opfid}) and (\ref{eqn:ConcNegSwitchEff}) that the entanglement extracted in both of these cases is 
useful in quantum teleportation. 

\begin{figure} 
\newlength{\widfig} 
\newlength{\widint} 
\setlength{\widfig}{18em} 
\setlength{\widint}{10em} 
\begin{center} 
\begin{minipage}{\textwidth} 
\begin{minipage}{\widfig} 
\begin{center} 
\includegraphics[scale=0.8,keepaspectratio]{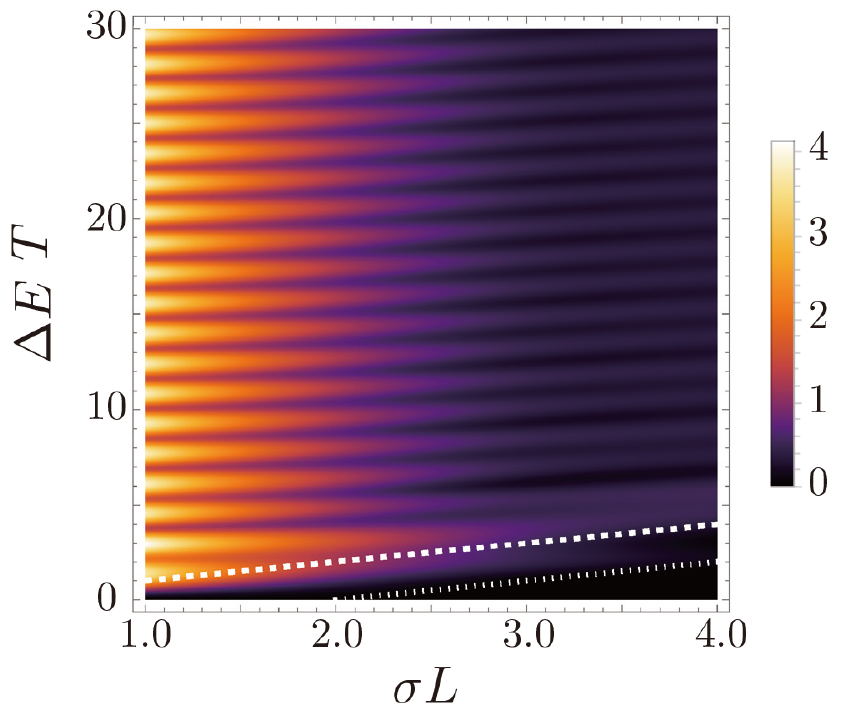} 
(a) 
\end{center} 
\end{minipage} 
\hspace*{\widint} 
\begin{minipage}{\widfig} 
\includegraphics[scale=0.8,keepaspectratio]{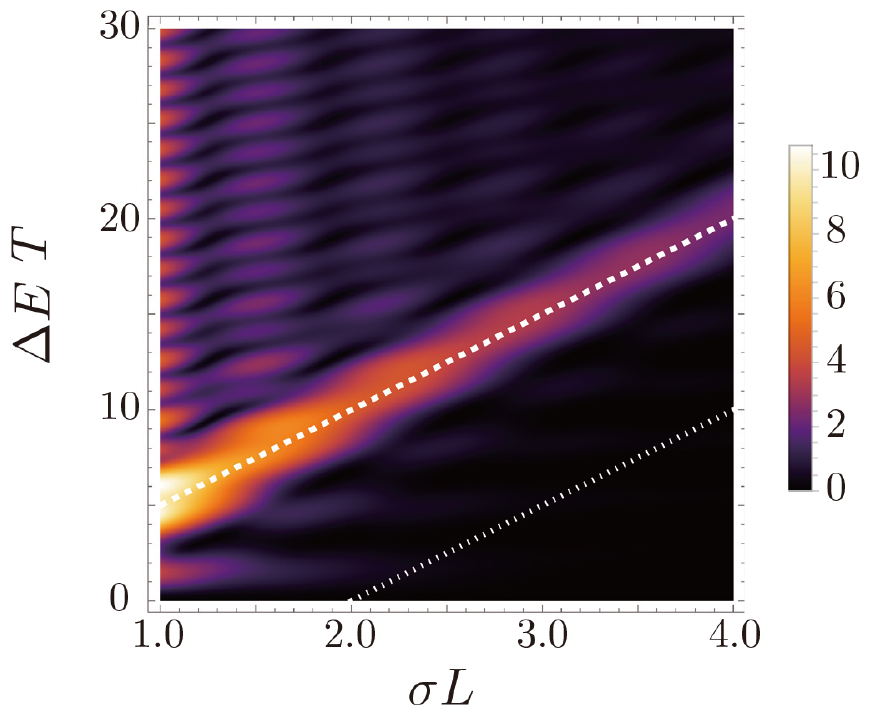} 
(b)
\end{minipage} 
\end{minipage} 
\vskip 7ex
\begin{minipage}{\textwidth} 
\begin{minipage}{\widfig} 
\includegraphics[scale=0.8,keepaspectratio]{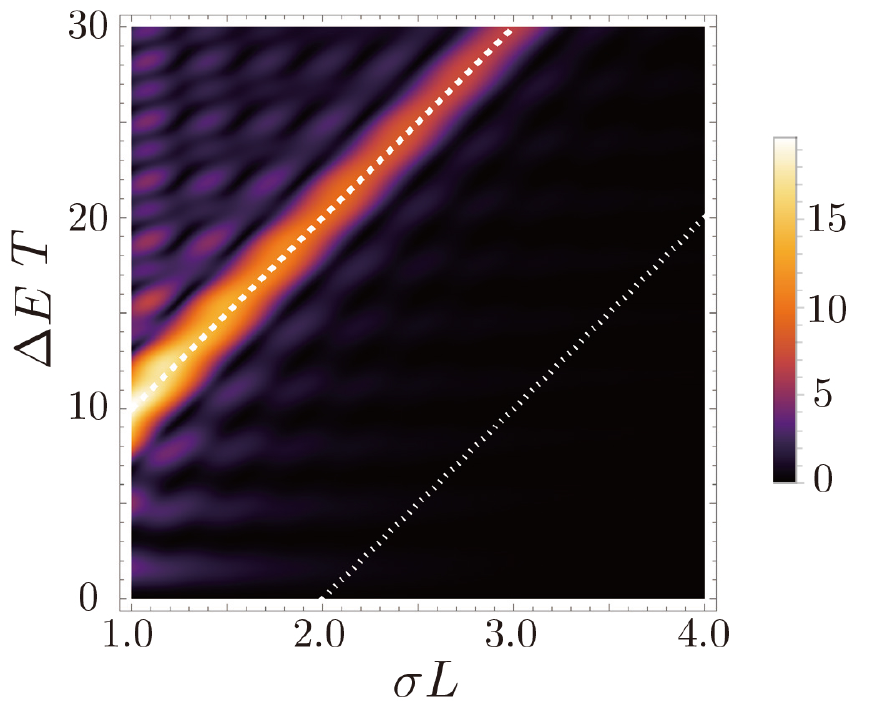} 
(c)
\end{minipage} 
\hspace*{\widint} 
\begin{minipage}{\widfig} 
\includegraphics[scale=0.8,keepaspectratio]{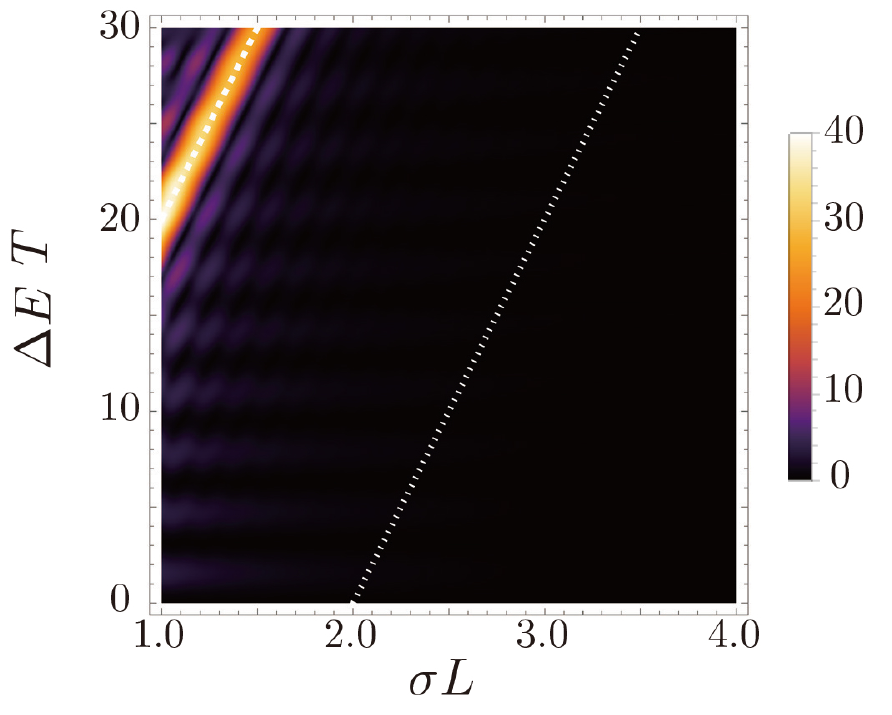} 
(d)
\end{minipage} 
\end{minipage} 
\end{center} 
\caption{The behavior of $\left| \mathcal{I}_E \right| - \mathcal{I}_I$ multiplied by $e^{\pi \, \Delta E / \sigma}$,  
(a) for $\Delta E / \sigma =2$, (b) for $\Delta E / \sigma = 10$, (c) for $\Delta E / \sigma = 20$, and (d) for $\Delta E / \sigma = 40$. 
The horizontal axis is the distance $L$ between the detectors normalized by $\sigma$, and the vertical axis is half the effective interaction time $T$ normalized by $\Delta E$. The region of $\sigma L < 1$ is not shown because too large a magnitude near $\sigma L = 0$ obscures the detailed structure elsewhere.   
L.E. and L.T. (see the main text) are denoted by a while dashed line and  
a white dotted line, respectively.}  
\label{fig:Entanglement} 
\end{figure}

However, in the case of adiabatic switching $\Delta E / \sigma \gg 1$, the maximal 
extraction of entanglement occurs around $L = 2 T$,  
which we expediently call the lightcone within the effective interaction time (L.E.). (As we described above, $2 T$ is the effective interaction time.)   
To see this, we depict in Fig. \ref{fig:Entanglement} the behavior of $\left| \mathcal{I}_E \right| - \mathcal{I}_I$, multiplied by $e^{\pi \, \Delta E / \sigma}$ so that the magnitude is not too small, on the $\sigma L - \Delta E \, T$ plane. 
Although we employ Eqs. (\ref{eqn:CalIIFinApp}) and (\ref{eqn:CalIEWithSwitch}) in the case where $\Delta E / \sigma$ 
is small, we need to resort to the approximate forms (\ref{eqn:CalIAdiabaitc}) and (\ref{eqn:CalEAdiabatic}) when $\Delta E / \sigma$ is large, due to the apparent numerical divergences in each of the Hurwitz-Lerch zeta functions that are analytically found to cancel among them. In Fig. \ref{fig:Entanglement}, along with the line of L.E. described by $\Delta E \, T = \left( \Delta E / 2 \sigma \right) \sigma L$, we plot what we call the lightcone within the total interaction time $2 ( T + 1 / \sigma )$ (L.T.), which is defined by the line $\Delta E \, T = \left( \Delta E / 2 \sigma \right) \left( \sigma L - 2 \right)$. The entanglement in the region between L.E. and L.T. is considered 
as arising from the disturbance due to switching on and off of the Unruh-DeWitt detectors. 

We see from Fig. \ref{fig:Entanglement} that there exist two periods generally, one along the $\Delta E \, T$ axis (as we see also from Eq, (\ref{eqn:EntAdiabiticLimit})),  
and the other along the normal direction to L.E., which is normal to L.T. also. As we decrease $\Delta E / \sigma$, 
L.E. and L.T. get tilted horizontally, and then the two periods become almost degenerate. 
We see also that the magnitude of $\left| \mathcal{I}_E \right| - \mathcal{I}_I$ gets flatter as $\Delta E / \sigma$ becomes small, which is in accord with the above argument on the sudden switching limit. 

On the other hand, for large values of $\Delta E / \sigma$,  the periodic behavior in the two different directions manifests itself. In particular, the maximum of the amount of the extracted entanglement is found to occur on L.E., and the oscillation in the normal direction to L.E. is drastically damped away from L.E.. 
Although the region between L.E. and L.T. gets wider and hence one can extract entanglement due to switching effects, 
it is quite small compared with the entanglement extracted due to causal propagation of the quantum field during 
the effective interaction time, i.e., on and inside L.E.. The entanglement extraction outside L.T. is possible, as we have seen from Eq. (\ref{eqn:EntAdiabiticLimit}), but even smaller. 
This implies that entanglement is extracted essentially due to causal propagation of the quantum field when 
we implement the switching of the Unruh-DeWitt detectors adiabatically, and it 
is consistent with the analysis in Sec. \ref{subsec:WithoutSWT}, where we have seen that the only contribution 
comes from the retarded Green function in the case of the implicit adiabatic switching, while it actually vanishes 
due to the energy conservation for the comoving case. 
Indeed, as we see from Eq. (\ref{eqn:CalEAdiabatic}), $\mathcal{I}_E$ vanishes in the limit of $T \rightarrow \infty$, 
because the first term falls off as $T^{-2}$, and the rest terms oscillate infinitely rapidly in the same manner as the delta function arises in Eq. (\ref{eqn:EntglBothAtRest}), i.e., when those are considered as the distribution of the variable $\Delta E$. 
Since $\mathcal{I}_I $ in Eq. (\ref{eqn:CalIAdiabaitc}) contains the term that does not vanish in the limit $T \rightarrow \infty$, we see that by explicitly considering the adiabatic switching effect and taking the limit of infinitely long exposure of the Unruh-DeWitt detectors to the interaction with the quantum field, entanglement is not generated between the comoving detectors, as the analysis in Sec. \ref{subsec:WithoutSWT} for the implicit adiabatic switching.

\section{Conclusion and Discussion} 
\label{sec:discussion} 

We considered in this paper two two-level Unruh-DeWitt detectors, as a pair of qubits, with an arbitrary monopole coupling to a neutral massless quantum scalar field in an arbitrary four-dimensional spacetime, 
and analyzed entanglement generated between the Unruh-DeWitt detectors from a vacuum. We first derived the general form of the reduced density matrix of the two Unruh-DeWitt detectors 
in the perturbation theory, for arbitrary worldlines of the detectors and arbitrary switching functions. 

We then considered the entanglement measures. Although we have not obtained evidences for usability of the entanglement from the analyses of the bounds on the distillable entanglement or the 
Bell-CHSH inequality, we did find that the single copy of the entangled pair of the Unruh-DeWitt detectors alone serves as the resource for quantum teleportation. 
More precisely, the optimal fidelity of the standard teleportation exceeds the classical value whenever a non-vanishing value of the concurrence (and hence the negativity) of the entanglement between symmetric ($\mathcal{P}_A = \mathcal{P}_B$) Unruh-DeWitt detectors arises from the first condition \eqref{eqn:EntanglementCond1} for entanglement, which is found within the second order perturbation theory. It is worthwhile to emphasize that this result is valid for an arbitrary monopole coupling with arbitrary switching functions and for arbitrary worldlines of the detectors in an arbitrary four-dimensional spacetime, 
while its extension into higher dimensionality will be straightforward. 

In order to find whether and how entanglement is actually generated between the Unruh-DeWitt detectors,  
we then focused on inertial motions of the Unruh-DeWitt detectors in the Minkowski vacuum. 
When we assume the implicit adiabatic switching, 
we found that no entanglement is extracted when Alice and Bob are comoving inertially. 
This is interpreted as resulting from the energy conservation due to an infinite amount of interaction time (transition time) 
and the uncertainty relation between time and energy. 
On the other hand, if Alice and Bob are in a relative inertial motion, entanglement was found to be 
extracted and enhanced by the special relativistic effect, obeying the first condition \eqref{eqn:EntanglementCond1} for entanglement, unless the relative velocity is ultra-relativistic. Therefore, we found that one can perform quantum teleportation by using 
the entanglement extracted in this manner, without invoking many copies of the entangled pair or preparing  an entangled state initially. 
Bob's desperate run in a relativistic speed in the vacuum suffices! 

By assuming the form (\ref{eqn:SwitchingFuncTanh}) of the switching function $\chi_I(\tau_I)$, 
we considered explicitly the switching effects also for the case where Alice and Bob are comoving inertially. 
In the case of adiabatic switching $\Delta E / \sigma \gg 1$, in particular, we found that entanglement arises 
primarily from causal propagation of the quantum field, which validated the analyses in the case of 
the implicit adiabatic switching. 
However, we noted that entanglement generation between the Unruh-DeWitt detectors separated acausally is possible also. 
Although our form of the switching function has the exponential tail and hence the terminology ``acausal'' does not have a 
rigorous sense, 
the analysis on the fidelity of quantum teleportation in this paper applies also to the case where the detectors are located in causally disconnected regions, as in Ref. \cite{Reznik03}. Therefore, we see that quantum teleportation is possible even if Alice and Bob 
are separated acausally in the strict sense.

The results in this paper may shed light on the physical process behind the entanglement generation between 
Unruh-DeWitt detectors. 
By regarding the quantum field as the continuum limit of discretized particles connected with springs, 
one may consider the entanglement extraction from a vacuum as resulting from the entanglement between 
these particles. 
Although a vacuum in the second quantization is the continuum limit of the product state of the ground state of each normal mode, 
it may be regarded as an entangled state by considering the Hilbert space of each {\it particle}. 
However, the transformation between the position operators of the particles and the normal modes is time independent, 
and then this picture of entanglement does not seem to be compatible with our result 
that entanglement generated between the comoving Unruh-DeWitt detectors depends on the interaction time $T$, especially 
in the limit of $T \rightarrow \infty$. 
As a more plausible picture of entanglement extraction, it may be possible to consider that entanglement is 
transferred from vacuum fluctuation \cite{Franson08}, whose correlation length extends acausally as described by the explicit form of the Wightman function Eq. (\ref{eqn:WightmanGen}). 
Vacuum fluctuation is nothing but virtual processes of creation and annihilation of quanta, and its effect is suppressed when interaction lasts for infinite time, due to the uncertainty relation between time and energy, which also is understood from Eq. (\ref{eqn:WightmanGen}). This will be the reason why the Wightman function does not contribute to 
entanglement extraction in the case of the implicit adiabatic switching, 
and the entanglement extracted outside L.E. 
in the case of explicit adiabatic switching is small.  
From the same reason, the contribution from the retarded Green function vanishes in the comoving case, because 
infinitely long interaction suppresses the effect of the virtual processes and then leads to the energy conservation, as indicated by the appearance of the delta function 
in Eq. (\ref{eqn:EntglBothAtRest}). 
However, this mechanism of suppression does not work completely when the detectors are in a relative motion, because of the variable distance between the detectors, as in the Doppler effect, which gives the modified 
Bessel function, instead of the delta function. This will explain why entanglement is extracted and enhanced   
by the special relativistic effect when the detectors are in a relative inertial motion. 

The analyses in this paper will provide applications and extensions. 
It may be interesting to investigate the relation of the entanglement between the detectors in a relative motion to the mechanism that gives rise to the revival of entanglement after entanglement sudden death. 
It also seems valuable to extend the analyses 
to the case of accelerated observers. In particular, 
when Alice is at rest and Bob is uniformly accelerated with the magnitude of the acceleration $\kappa$, our preliminary calculation in the case of the Minkowski vacuum gives 
\begin{equation} 
\left| \mathcal{I}_E \right| 
= \left| \frac{e^{\pi \frac{\Delta E^{(B)}}{\kappa}}}{e^{2 \pi \frac{\Delta E^{(B)}}{\kappa}} - 1} 
\sinh \left( \frac{\pi}{2} \frac{\Delta E^{(B)}}{\kappa} - \frac{\Delta E^{(A)}}{\kappa} \right) \right| ,  
\label{eqn:CalIRindCalc} 
\end{equation} 
for the implicit adiabatic switching $\chi_I(\tau_I) \equiv 1$. Since $\mathcal{P}_A = 0$ in this case (arbitrarily small even in the case of 
the explicit adiabatic switching), we see from 
Eq. (\ref{eqn:EntanglementCond1}) that entanglement is extracted if Eq. (\ref{eqn:CalIRindCalc}) is non-vanishing. Although we have not arrived at complete understanding  
of this result, it is interesting to note that even when Bob's worldline is very close to the Rindler horizon, where 
$\kappa \rightarrow \infty$, entanglement is generated between the Unruh-DeWitt detectors. 
(In this case, massless quanta emitted from Bob reach Alice, even after Alice passes across Bob's event horizon, 
in contrast to the case of the $v_0 \rightarrow 1$ limit of relative inertial motion.)  
In a recent paper \cite{HerdersonHMSZ17}, the authors considered the case where both of the detectors are 
accelerated in the same direction in the B. T. Z. black hole spacetime. Comparison of Ref. \cite{HerdersonHMSZ17} 
and our preliminary result above may provide a clue to the information loss paradox. 

\vskip 2ex 

{\it Note added in proof.} 

After this paper was submitted, a paper by Ng et al. \cite{NgMM18} appeared, which also derived the expressions equivalent to 
Eqs. (\ref{eqn:CalEDefApp1}) and (\ref{eqn:CalEDefApp2}) for spatially extended detectors. 

\acknowledgments 
G. K. is grateful to Prof. K. Matsumoto for his useful discussion. 
This work was supported in part by JSPS KAKENHI Grant Number 17K18107 and 17K05451. 
\appendix 

\section{Explicit form of reduced density matrix} 
\label{sec:ExplicitMatrixElements} 

In this appendix, we describe briefly the computation of the elements 
$\bra{E_{n_A}^{(A)}} \bra{E_{n_B}^{(B)}} \rho_{AB} \cket{E_{\hat{n}_A}^{(A)}} \cket{E_{\hat{n}_B}^{(B)}}$ 
of the reduced density matrix (\ref{eqn:RedDensityMatrix}), and 
present their explicit forms. 

Expanding $e^{\ImUnit \mathcal{S}_{{\rm int}}}$ in Eq. (\ref{eqn:OutState}) to second order in $c$, 
the matrix elements of the reduced density matrix $\rho_{AB} \equiv \mathrm{Tr}_{\phi} \cket{\mathrm{out}} \bra{\mathrm{out}}$ 
are found to be written as  
\begin{align}  & 
\bra{E_{n_A}^{(A)}} \bra{E_{n_B}^{(B)}} \rho_{AB} \cket{E_{\hat{n}_A}^{(A)}} \cket{E_{\hat{n}_B}^{(B)}} 
\notag \\ & 
= \Big[ \delta^{n_A}_0 \, \delta^{n_B}_0 \: \delta^{\hat{n}_A}_0 \delta^{\hat{n}_B}_0 
+ c^2 \, \left\{ 
R^{(1)}_{n_A , n_B , \hat{n}_A , \hat{n}_B} 
+ R^{(2)}_{n_A , n_B , \hat{n}_A , \hat{n}_B} 
+ R^{(2) \, *}_{\hat{n}_A , \hat{n}_B , n_A , n_B} \right\} \Big] + O(c^4),  
\end{align} 
where 
\begin{align} & 
R^{(1)}_{n_A , n_B , \hat{n}_A , \hat{n}_B} 
= \delta^{n_B}_0 \: \delta^{\hat{n}_B}_0 \, \bra{E_{n_A}^{(A)}} \, m_A(0) \, \cket{E_0^{(A)}} \, \bra{E_{\hat{n}_A}^{(A)}} \, m_A(0) \, \cket{E_0^{(A)}}^\dagger \, 
\notag \\ & 
\times 
\int_{- \infty}^{\infty} d \tau'_A \; \int_{- \infty}^{\infty} d \tau_A \; \chi_A(\tau'_A) \: \chi_A(\tau_A) \: 
e^{\ImUnit \left( E_{n_A}^{(A)} - E_0^{(A)} \right) \tau_A} e^{- \ImUnit \left( E_{\hat{n}_A}^{(A)} - E_0^{(A)} \right) \tau'_A} 
\bra{0} \phi(\bar{x}'_A) \phi(\bar{x}_A) \, \cket{0} 
\notag \\ & \
+ \delta^{n_A}_0 \: \delta^{\hat{n}_B}_0  \, \bra{E_{n_B}^{(B)}} \: \, m_B(0) \, \cket{E_0^{(B)}} \, \bra{E_{\hat{n}_A}^{(A)}} \, m_A(0) \, \cket{E_0^{(A)}}^\dagger \, 
\notag \\ & 
\times 
\int_{- \infty}^{\infty} d \tau'_A \; \int_{- \infty}^{\infty} d \tau_B  \; \chi_A(\tau'_A) \: \chi_B(\tau_B) \: 
e^{\ImUnit \left( E_{n_B}^{(B)} - E_0^{(B)} \right) \tau_B} e^{- \ImUnit \left( E_{\hat{n}_A}^{(A)} - E_0^{(A)} \right) \tau'_A} 
\bra{0} \phi(\bar{x}'_A) \phi(\bar{x}_B) \, \cket{0} 
\notag \\ & 
+ \delta^{n_B}_0 \: \delta^{\hat{n}_A}_0 \, \bra{E_{n_A}^{(A)}} \, m_A(0) \, \cket{E_0^{(A)}} \, \bra{E_{\hat{n}_B}^{(B)}} \: \, m_B(0) \, \cket{E_0^{(B)}}^\dagger \, 
\notag \\ & 
\times 
\int_{- \infty}^{\infty} d \tau'_B \; \int_{- \infty}^{\infty} d \tau_A \; \chi_B(\tau'_B) \: \chi_A(\tau_A) \: 
e^{\ImUnit \left( E_{n_A}^{(A)} - E_0^{(A)} \right) \tau_A} e^{- \ImUnit \left( E_{\hat{n}_B}^{(B)} - E_0^{(B)} \right) \tau'_B} 
\bra{0} \phi(\bar{x}'_B) \phi(\bar{x}_A) \, \cket{0} 
\notag \\ &  
+ \delta^{n_A}_0 \: \delta^{\hat{n}_A}_0 \, \bra{E_{n_B}^{(B)}} \: \, m_B(0) \, \cket{E_0^{(B)}} \, \bra{E_{\hat{n}_B}^{(B)}} \: \, m_B(0) \, \cket{E_0^{(B)}}^\dagger \, 
\notag \\ & 
\times 
\int_{- \infty}^{\infty} d \tau'_B \; \int_{- \infty}^{\infty} d \tau_B  \; \chi_B(\tau'_B) \: \chi_B(\tau_B) \: 
e^{\ImUnit \left( E_{n_B}^{(B)} - E_0^{(B)} \right) \tau_B} e^{- \ImUnit \left( E_{\hat{n}_B}^{(B)} - E_0^{(B)} \right) \tau'_B} 
\bra{0} \phi(\bar{x}'_B) \phi(\bar{x}_B) \, \cket{0} 
\label{eqn:CalRFComps}
\end{align} 
and 
\begin{align} & 
R^{(2)}_{n_A , n_B , \hat{n}_A , \hat{n}_B} 
= - \delta^{n_B}_0 \: \delta^{\hat{n}_A}_0 \: \delta^{\hat{n}_B}_0 
\sum_{k} \bra{E_{n_A}^{(A)}} m_A(0) \, \cket{E_k^{(A)}} \bra{E_k^{(A)}} m_A(0) \cket{E_0^{(A)}} \: 
\notag \\ & \times \; 
\int_{- \infty}^{\infty} d \tau_A \, \int_{- \infty}^{\infty} d \tau'_A \: \chi_A(\tau_A) \, \chi_A(\tau'_A) \, 
\Theta(\tau_A - \tau'_A) \: e^{\ImUnit \left( E_{n_A}^{(A)} - E_k^{(A)} \right) \tau_A} 
e^{\ImUnit \left( E_k^{(A)} - E_0^{(A)} \right) \tau'_A} \:  
\bra{0} T \, \phi(\bar{x}_A) \phi(\bar{x}'_A) \cket{0} 
\notag \\ & 
- \delta^{\hat{n}_A}_0 \: \delta^{\hat{n}_B}_0 
\bra{E_{n_B}^{(B)}} m_B(0) \cket{E_0^{(B)}} \, \bra{E_{n_A}^{(A)}} m_A(0) \cket{E_0^{(A)}} \, 
\notag \\ & \times \; 
\int_{- \infty}^{\infty} d \tau_B \, \int_{- \infty}^{\infty} d \tau'_A \:  \chi_B(\tau_B) \, \chi_A(\tau'_A) \, 
e^{\ImUnit \left( E_{n_B}^{(B)} - E_0^{(B)} \right) \tau_B} e^{\ImUnit \left( E_{n_A}^{(A)} - E_0^{(A)} \right) \tau'_A} \: 
\bra{0} T \, \phi(\bar{x}_B) \phi(\bar{x}'_A) \cket{0} 
\notag \\ &  
- \delta^{n_A}_0 \: \delta^{\hat{n}_A}_0 \: \delta^{\hat{n}_B}_0 
\sum_{k} \bra{E_{n_B}^{(B)}} m_B(0) \, \cket{E_k^{(B)}} \bra{E_k^{(B)}} m_B(0) \cket{E_0^{(B)}} \: 
\notag \\ & \times \; 
\int_{- \infty}^{\infty} d \tau_B \, \int_{- \infty}^{\infty} d \tau'_B \:  \chi_B(\tau_B) \, \chi_B(\tau'_B) \, 
\Theta(\tau_B - \tau'_B) \: e^{\ImUnit \left( E_{n_B}^{(B)} - E_k^{(B)} \right) \tau_B} 
e^{\ImUnit \left( E_k^{(B)} - E_0^{(B)} \right) \tau'_B} \: \bra{0} T \, \phi(\bar{x}_B) \phi(\bar{x}'_B) \cket{0} 
\label{eqn:CalRSComps} 
\end{align} 
Here we employ the causality relations Eq. (\ref{eqn:Causality}) and the properties of the Feynman propagator $G_F(x , x')$ and the Wightman function $G_W(x , x')$, 
\begin{align} & 
G_F(x , x') = - \, \ImUnit \, \big[ \Theta(t - t') \, G_W(x, x') + \Theta(t' - t) \, G_W(x' , x) \big] , 
\\ & 
G_W(x , x') - \ImUnit \, G_F(x , x') = - \, \ImUnit \, G_R(x' , x) , 
\label{eqn:RelRetFeyWight} \\ & 
G_F(x , x') = G_F(x' , x) , \quad 
G^*_R(x , x') = G_R(x , x') , \quad 
G_W^*(x , x') = G_W(x' , x) , 
\end{align} 
which are derived from their definitions Eqs. (\ref{eqn:WightmanNSDefGen}) and (\ref{eqn:FeynmanNSDefGen}), 
along with 
\begin{equation}
G_R(x, x') \equiv - \, \ImUnit \, \Theta(t - t') \, \bra{0} \left[ \phi(x) , \phi(x') \right] \cket{0}  
= \ImUnit \, \Theta(t - t') \, \big[ G_W(x' , x) - G_W(x , x') \big] , 
\label{eqn:RetardedNSDelGen} 
\end{equation}
One finds that the reduced density matrix $\rho_{AB}$ takes the form of Eq. (\ref{eqn:RedDensityMatrix}) 
with the non-vanishing elements given by 
\begin{align} & 
\mathcal{P}_A \equiv 
\left| \bra{E_{1}^{(A)}} \, m_A(0) \, \cket{E_0^{(A)}} \right|^2 \, 
\int_{- \infty}^{\infty} d \tau'_A \; \int_{- \infty}^{\infty} d \tau_A \; \chi_A(\tau'_A) \: \chi_A(\tau_A) \: 
e^{\ImUnit \, \Delta E^{(A)} \left( \tau_A-  \tau'_A \right)} \: G_W(\bar{x}'_A , \bar{x}_A) , 
\\ & 
\mathcal{P}_B \equiv 
\left| \bra{E_{1}^{(B)}} \: \, m_B(0) \, \cket{E_0^{(B)}} \right|^2 \, 
\int_{- \infty}^{\infty} d \tau'_B \; \int_{- \infty}^{\infty} d \tau_B  \; \chi_B(\tau'_B) \: \chi_B(\tau_B) \: 
e^{\ImUnit \, \Delta E^{(B)} \left( \tau_B - \tau'_B \right)} \: G_W(\bar{x}'_B , \bar{x}_B) ,  
\\ & 
\mathcal{E} \equiv
- \:  \bra{E_{1}^{(B)}} m_B(0) \cket{E_0^{(B)}} \, \bra{E_{1}^{(A)}} m_A(0) \cket{E_0^{(A)}} \, 
\notag \\ & \qquad \times \; 
\int_{- \infty}^{\infty} d \tau_B \, \int_{- \infty}^{\infty} d \tau'_A \:  \chi_B(\tau_B) \, \chi_A(\tau'_A) \, 
e^{\ImUnit \, \Delta E^{(B)} \tau_B} e^{\ImUnit \, \Delta E^{(A)} \tau'_A} \: \ImUnit \: G_F(\bar{x}_B , \bar{x}'_A)   
\label{eqn:CalEDefApp1} \\ & 
= - \:  \bra{E_{1}^{(B)}} m_B(0) \cket{E_0^{(B)}} \, \bra{E_{1}^{(A)}} m_A(0) \cket{E_0^{(A)}} \, 
\notag \\ & \qquad \times \; 
\int_{- \infty}^{\infty} d \tau_B \, \int_{- \infty}^{\infty} d \tau'_A \:  \chi_B(\tau_B) \, \chi_A(\tau'_A) \, 
e^{\ImUnit \, \Delta E^{(B)} \tau_B} e^{\ImUnit \, \Delta E^{(A)} \tau'_A} \: 
\Big[ G_W(\bar{x}_B , \bar{x}'_A) + \ImUnit \, G_R(\bar{x}'_A , \bar{x}_B) \Big] , 
\label{eqn:CalEDefApp2} 
\end{align} 
\begin{align} & 
\mathcal{P}_{AB} \equiv 
\bra{E_{1}^{(A)}} \, m_A(0) \, \cket{E_0^{(A)}} \, \bra{E_{1}^{(B)}} \: \, m_B(0) \, \cket{E_0^{(B)}}^\dagger \, 
\notag \\ & \qquad 
\times 
\int_{- \infty}^{\infty} d \tau'_B \; \int_{- \infty}^{\infty} d \tau_A \; \chi_B(\tau'_B) \: \chi_A(\tau_A) \: 
e^{\ImUnit \, \Delta E^{(A)} \tau_A - \ImUnit \, \Delta E_{1}^{(B)} \tau'_B} \: G_W(\bar{x}'_B , \bar{x}_A) , 
\label{eqn:CalPABDefApp} \\ & 
\mathcal{W}_A \equiv   
- \bra{E_1^{(A)}} m_A(0) \cket{E_0^{(A)}} \, 
\left[ \bra{E_{1}^{(A)}} m_A(0) \, \cket{E_1^{(A)}} - \bra{E_{0}^{(A)}} m_A(0) \, \cket{E_0^{(A)}} \right] \, 
\notag \\ & \qquad \times \; 
\int_{- \infty}^{\infty} d \tau_A \, \int_{- \infty}^{\infty} d \tau'_A \: \chi_A(\tau_A) \, \chi_A(\tau'_A) \, 
\Theta(\tau'_A - \tau_A) \: e^{\ImUnit \, \Delta E^{(A)} \tau_A} \: G_W(\bar{x}'_A , \bar{x}_A) 
\notag \\ & 
- \: \ImUnit \: \bra{E_{1}^{(A)}} \, m_A(0) \, \cket{E_0^{(A)}} \, \bra{E_{0}^{(B)}} \: \, m_B(0) \, \cket{E_0^{(B)}} \, 
\int_{- \infty}^{\infty} d \tau'_B \; \int_{- \infty}^{\infty} d \tau_A \; \chi_B(\tau'_B) \: \chi_A(\tau_A) \: 
e^{\ImUnit \, \Delta E^{(A)} \tau_A} \: G_R(\bar{x}_A , \bar{x}'_B) 
\notag \\  & 
- \: \ImUnit \:  \bra{E_{1}^{(A)}} \, m_A(0) \, \cket{E_0^{(A)}} \, \bra{E_{0}^{(A)}} \, m_A(0) \, \cket{E_0^{(A)}} \, 
\int_{- \infty}^{\infty} d \tau'_A \; \int_{- \infty}^{\infty} d \tau_A \; \chi_A(\tau'_A) \: \chi_A(\tau_A) \: 
e^{\ImUnit \, \Delta E^{(A)} \tau_A} \: G_R(\bar{x}_A , \bar{x}'_A) , \label{eq:WA} 
\\ & 
\mathcal{W}_B \equiv 
- \; \bra{E_1^{(B)}} m_B(0) \cket{E_0^{(B)}} \, 
\Big[ \bra{E_{1}^{(B)}} m_B(0) \, \cket{E_1^{(B)}} - \bra{E_{0}^{(B)}} m_B(0) \, \cket{E_0^{(B)}} \Big] 
\notag \\ & \times \; 
\int_{- \infty}^{\infty} d \tau_B \, \int_{- \infty}^{\infty} d \tau'_B \:  \chi_B(\tau_B) \, \chi_B(\tau'_B) \, 
\Theta(\tau'_B - \tau_B) \: e^{\ImUnit \, \Delta E^{(B)} \tau_B} \: G_W(\bar{x}'_B , \bar{x}_B)
\notag \\ & 
- \ImUnit \: \bra{E_{1}^{(B)}} \: \, m_B(0) \, \cket{E_0^{(B)}} \, \bra{E_{0}^{(A)}} \, m_A(0) \, \cket{E_0^{(A)}} \, 
\int_{- \infty}^{\infty} d \tau'_A \; \int_{- \infty}^{\infty} d \tau_B  \; \chi_A(\tau'_A) \: \chi_B(\tau_B) \: 
e^{\ImUnit \, \Delta E^{(B)} \tau_B} \: 
G_R(\bar{x}_B , \bar{x}'_A) 
\notag \\ 
- &  \; \ImUnit \: \bra{E_{1}^{(B)}} \: \, m_B(0) \, \cket{E_0^{(B)}} \, \bra{E_{0}^{(B)}} \: \, m_B(0) \, \cket{E_0^{(B)}} \, 
\int_{- \infty}^{\infty} d \tau'_B \; \int_{- \infty}^{\infty} d \tau_B  \; \chi_B(\tau'_B) \: \chi_B(\tau_B) \: 
e^{\ImUnit \, \Delta E^{(B)} \tau_B} \: 
G_R(\bar{x}_B , \bar{x}'_B). \label{eq:WB}  
\end{align} 

\section{Eigenvalues of matrices} 
\label{sec:Eigenvalues} 

In this appendix, we outline the derivations of the eigenvalues of the matrices in the main text. 
In particular, we shall see that $\mathcal{W}_I$ does not appear in the leading contributions to 
the eigenvalues. 

The density matrix $\rho_{AB}$ given in Eq. (\ref{eqn:RedDensityMatrix}) and its partial transpose $\rho_{AB}^{T_A}$ 
in Eq. (\ref{eqn:PartTransRho}) take the same form, 
\begin{equation} 
P = \begin{pmatrix} 0 & 0 & 0 & c^2 \, \alpha \\ 
0 & c^2 \, A & c^2 \, \beta & c^2 \, \sigma \\ 
0 & c^2 \, \beta^* & c^2 \, B & c^2 \, \kappa \\ 
c^2 \, \alpha^* & c^2 \, \sigma^* & c^2 \, \kappa^* & 1 - c^2 ( A + B ) \end{pmatrix} 
+ c^4  \,\begin{pmatrix} Q_{11} & Q_{12} & Q_{13} &Q_{14} \\ 
Q_{21} & Q_{22} & Q_{23} &Q_{24} \\ Q_{31} & Q_{32} & Q_{33} &Q_{34} \\ 
Q_{41} & Q_{42} & Q_{43} & Q_{44} \end{pmatrix} ,  
\end{equation} 
where $A$ and $B$ are real, $\alpha$, $\beta$, $\sigma$, and $\kappa$ are complex constants, 
and $Q_{i j} = Q_{ji}^*$ is conditioned to satisfy $Q_{11} + Q_{22} + Q_{33} + Q_{44} = 0$. 
Thus, their eigenvalues are derived in a single stroke. A straightforward calculation shows that the eigenvalue equation is given by 
\begin{align} 
0 = \det ( P  - \lambda \mathop{\mathbb{I}}\nolimits ) 
= & \; \lambda^4 - \lambda^3 + \left[ c^2 \left( A + B \right) + O(c^4) \right] \lambda^2 
- \left[ c^4 \left( A \, B - \left| \beta \right|^2 \right) + O(c^6) \right] \lambda 
\notag \\ & 
+ c^8 \left( A \, B - \left| \beta \right|^2 \right) \left( Q_{11} - \left| \alpha \right|^2 \right) + O(c^{10}) . 
\label{eqn:EigenEqRhoS} 
\end{align} 
Since $\sigma$ or $\kappa$ do not contribute to leading order in Eq. (\ref{eqn:EigenEqRhoS}), we expect that they 
will not appear in leading order of the eigenvalues, either. Indeed, Eq. (\ref{eqn:EigenEqRhoS}) is 
found to factorize as 
\begin{align} 
0 = \det ( P  - \lambda \mathop{\mathbb{I}}\nolimits ) 
= & \; \left\{ \lambda^2 - \left[ c^2 \left( A + B \right) + O(c^4) \right] \lambda 
+ c^4 \left( A B - \left| \beta \right|^2 \right) + O(c^6) \right\} 
\notag \\ & \times 
\left\{ \lambda^2 - \left[ 1 - c^2 \left( A + B \right) + O(c^4) \right] \lambda 
+ c^4 \left( Q_{11} - \left| \alpha \right|^2 \right) + O(c^6) \right\} , 
\notag 
\end{align} 
and thus the leading terms of the eigenvalues of $P$ are derived as 
\begin{equation}
1 + O(c^2) , \quad \frac{c^2}{2} \left[ A + B \pm \sqrt{\left( A - B \right)^2 + 4 \left| \beta \right|^2} \right] 
+ O(c^4) , \quad c^4 \left( Q_{11} - \left| \alpha \right|^2 \right) + O(c^6) . 
\label{eqn:EigenValuesRhoS} 
\end{equation}
We now set as $A = \mathcal{P}_A$, $B = \mathcal{P}_B$, and $Q_{11} = \mathcal{X}$. 
When we further set as $\alpha = \mathcal{E}$, $\beta = \mathcal{P}_{AB}$, $\sigma = \mathcal{W}_A$, 
and $\kappa = \mathcal{W}_B$, 
we obtain Eq. (\ref{eqn:EigenVRho}), while $\alpha = \mathcal{P}^*_{AB}$, $\beta = \mathcal{E}^*$, 
$\sigma = \mathcal{W}^*_A$, and $\kappa = \mathcal{W}_B$ give Eq. (\ref{eqn:EigenVPartTransRho}). 

Similarly, the eigenvalue equation of $\rho_{AB} \, \tilde{\rho}_{AB}$ computed from 
Eqs. (\ref{eqn:RedDensityMatrix}) and (\ref{eqn:TildeRhoDef}) is 
found to be given as 
\begin{align} &
0 = \det ( \rho_{AB} \, \tilde{\rho}_{AB}  - \lambda \mathop{\mathbb{I}}\nolimits ) 
= \lambda^4 - \left[ 2 \, c^4 \left( \mathcal{X} + \left| \mathcal{E} \right|^2 + \mathcal{P}_A \mathcal{P}_B 
+ \left| \mathcal{P}_{AB} \right|^2 \right) + O(c^6) \right] \lambda^3 
\notag \\ & 
+ \left[ c^8 \left\{ \left( \mathcal{X} - \left| \mathcal{E} \right|^2 \right)^2 
+ \left( \mathcal{P}_A \mathcal{P}_B - \left| \mathcal{P}_{AB} \right|^2 \right)^2 
+ 4 \left( \mathcal{X} + \left| \mathcal{E} \right|^2 \right) \left( \mathcal{P}_A \mathcal{P}_B + \left| \mathcal{P}_{AB} \right|^2 \right) \right\} + O(c^{10}) \right] \lambda^2 
\notag \\ & 
- \left[ 2 \, c^{12} \left\{ \left( \mathcal{X} + \left| \mathcal{E} \right|^2 \right) 
\left( \mathcal{P}_A \mathcal{P}_B - \left| \mathcal{P}_{AB} \right|^2 \right)^2 
+ \left( \mathcal{X} - \left| \mathcal{E} \right|^2 \right)^2 \left( \mathcal{P}_A \mathcal{P}_B + \left| \mathcal{P}_{AB} \right|^2 \right) \right\} + O(c^{14}) \right] \lambda 
\notag \\ & 
+ c^{16} \left( \mathcal{X} - \left| \mathcal{E} \right|^2 \right)^2 \left( \mathcal{P}_A \mathcal{P}_B - \left| \mathcal{P}_{AB} \right|^2 \right)^2 + O(c^{18}) , 
\end{align} 
which is found to be factorized as 
\begin{align} & 
\det ( \rho_{AB} \, \tilde{\rho}_{AB}  - \lambda \mathop{\mathbb{I}}\nolimits ) 
= \Big[ \lambda^2 - \left\{ 2 \, c^4 \left( \mathcal{X} + \left| \mathcal{E} \right|^2 \right) + O(c^6) \right\} \lambda 
+ c^8 \left( \mathcal{X} - \left| \mathcal{E} \right|^2 \right)^2 + O(c^{10}) \Big] 
\notag \\ & \times 
\Big[ \lambda^2 - \left\{ 2 \, c^4 \left( \mathcal{P}_A \mathcal{P}_B + \left| \mathcal{P}_{AB} \right|^2 \right) 
+ O(c^6) \right\} \lambda + c^8 \left( \mathcal{P}_A \mathcal{P}_B - \left| \mathcal{P}_{AB} \right|^2 \right)^2 + O(c^{10}) \Big] . 
\label{eqn:EigenValueEqRhoTIlRho} 
\end{align} 
Then, we see that the square roots of the eigenvalues of $\rho_{AB} \, \tilde{\rho}_{AB}$ are given as Eq. (\ref{eqnEigenValueRhoTildeRho}). 

The matrix $T_{\rho_{AB}}$ defined by Eq. (\ref{eqn:MtrxTDef}) is computed from Eq. (\ref{eqn:RedDensityMatrix}) as 
\begin{equation}
T_{\rho_{AB}} = \begin{pmatrix} 
c^2 \left( \mathcal{E} + \mathcal{E}^* + \mathcal{P}_{AB} + \mathcal{P}_{AB}^* \right) & 
\ImUnit \: c^2 \left( \mathcal{E} - \mathcal{E}^* - \mathcal{P}_{AB} + \mathcal{P}_{AB}^* \right) & 
- \: c^2 \left( \mathcal{W}_A + \mathcal{W}_A^* \right) \\ 
\ImUnit \: c^2 \left( \mathcal{E} - \mathcal{E}^* + \mathcal{P}_{AB} - \mathcal{P}_{AB}^* \right) & 
c^2 \left( - \mathcal{E} - \mathcal{E}^* + \mathcal{P}_{AB} + \mathcal{P}_{AB}^* \right) & 
\ImUnit \: c^2 \left( - \mathcal{W}_A + \mathcal{W}_A^* \right) \\ 
- c^2 \left( \mathcal{W}_B + \mathcal{W}_B^* \right) & 
\ImUnit \: c^2 \left( - \mathcal{W}_B + \mathcal{W}_B^* \right) & 
1 - 2 \: c^2 \big( \mathcal{P}_A + \mathcal{P}_B \big) \end{pmatrix} + O(c^4) , 
\end{equation}
and then the eigenvalue equation of $U_{\rho_{AB}}$ is derived and factorized as 
\begin{align} 
0 = \det \left( U_{\rho_{AB}} - \lambda \mathop{\mathbb{I}}\nolimits \right) 
=  & \; - \lambda^3 + \left[ 1 - 4 \,c^2 \left( \mathcal{P}_A + \mathcal{P}_B \right) + O(c^4) \right] \lambda^2 
- \left[ 8 \, c^4 \left( \left| \mathcal{E} \right|^2 + \left| \mathcal{P}_{AB} \right|^2 \right) + O(c^6) \right] \lambda 
\notag \\ & 
+ 16 \, c^8 \left( \left| \mathcal{E} \right|^2 - \left| \mathcal{P}_{AB} \right|^2 \right)^2 + O(c^{10}) 
\notag \\ 
= & \;  - \Big\{ \lambda - \left[ 1 - 4 \,c^2 \left( \mathcal{P}_A + \mathcal{P}_B \right) + O(c^4) \right] \Big\} 
\notag \\ & \times 
\Big\{ \lambda^2 - \left[ 8 \, c^4 \left( \left| \mathcal{E} \right|^2 + \left| \mathcal{P}_{AB} \right|^2 \right) + O(c^6) \right] \lambda 
+ 16 \, c^8 \left( \left| \mathcal{E} \right|^2 - \left| \mathcal{P}_{AB} \right|^2 \right)^2 + O(c^{10}) \Big\} , 
\label{eqn:EigenValueEqU} 
\end{align} 
where $U_{\rho_{AB}} = T_{\rho_{AB}}^T \, T_{\rho_{AB}}$ as defined in Eq. (\ref{eq:U}). 
Then one finds that the eigenvalues of $U_{\rho_{AB}}$ are given by Eq. (\ref{eq:eig}).

\vskip 2cm
\baselineskip .2in

\end{document}